\documentclass[aip,reprint,twocolumn,showpacs,showkeys,amsmath,amssymb]{revtex4-1}

\usepackage{graphicx}
\usepackage{dcolumn}
\usepackage{bm}
\usepackage{color}
\usepackage{natbib}
\usepackage{hyperref} 

\begin{document}

\title{Normal force controlled rheology applied to agar gelation} 

\author{Bosi Mao}
\email[]{mao@crpp-bordeaux.cnrs.fr}
\affiliation{Centre de Recherche Paul Pascal, CNRS UPR~8641 - Universit\'e de Bordeaux, 115 avenue Dr. Schweitzer, 33600 Pessac, France.}

\author{Thibaut Divoux}
\email[]{divoux@crpp-bordeaux.cnrs.fr}
\affiliation{Centre de Recherche Paul Pascal, CNRS UPR~8641 - Universit\'e de Bordeaux, 115 avenue Dr. Schweitzer, 33600 Pessac, France.}

\author{Patrick Snabre}
\email[]{snabre@crpp-bordeaux.cnrs.fr}
\affiliation{Centre de Recherche Paul Pascal, CNRS UPR~8641 - Universit\'e de Bordeaux, 115 avenue Dr. Schweitzer, 33600 Pessac, France.}

\date{\today}

\begin{abstract}
A wide range of thermoreversible gels are prepared by cooling down to ambient temperature hot aqueous polymer solutions. During the sol-gel transition, such materials may experience a volume contraction which is traditionally overlooked as rheological measurements are usually performed in geometries of constant volume. In this article, we revisit the formation of 1.5\% wt. agar gels through a series of benchmark rheological experiments performed with a plate-plate geometry. We demonstrate on that particular gel of polysaccharides that the contraction associated withe the sol/gel transition cannot be neglected. Indeed, imposing a constant gap width during the gelation results in the strain hardening of the sample, as evidenced by the large negative normal force that develops. Such hardening leads to the slow drift in time of the gel elastic modulus $G'$ towards ever larger values, and thus to an erroneous estimate of $G'$. As an alternative, we show that imposing a constant normal force equals to zero during the gelation, instead of a constant gap width, suppresses the hardening as the decrease of the gap compensates for the sample contraction. As such, imposing a zero normal force is a more reliable way to measure the linear properties of agar gels, which we prove to work equally well with rough and smooth boundary conditions. Using normal force controlled rheology, we then investigate the impact of thermal history on 1.5\% wt. agar gels. We show that neither the value of the cooling rate, nor the introduction of a constant temperature stage during the cooling process influence the gel elastic properties. Instead, $G'$ only depends on the terminal temperature reached at the end of the cooling ramp, as confirmed by direct imaging of the gel microstructure by cryoelectron microscopy. Finally, we also discuss two subtle artifacts associated with the use of duralumin plates that may interfere with the rheological measurements of agar gelation. We show that ($i$) the corrosion of duralumin by the aqueous solution, and ($ii$) the slow migration of the oil rim added around the sample to prevent evaporation, may both lead separately to a premature and artificial growth of $G'$ that should not be misinterpreted as the formation of a pre-gel. The present work offers an extensive review of the technical difficulties associated with the rheology of hydrogels and paves the way for a systematic use of normal force controlled rheology to monitor non-isochoric processes.
 \end{abstract}

\pacs{}

\maketitle 

\section{Introduction}

Bulk rheology is a powerful tool to monitor the linear properties of complex fluids such as polymer and surfactant solutions, emulsions, colloidal suspensions, gels, etc. \cite{Mezger:2014}. Although most traditional rheology experiments are performed at constant temperature, several processes involve large temperature variations that make rheological measurements much more challenging. A textbook case of rheological experiments involving temperature variations, is the sol-gel transition. Popular examples include the jamming transition in suspensions of thermosensitive particles \cite{Senff:1999,Kapnistos:2000,Zhang:2009,Wang:2014b}, the precipitation of waxy crude oil \cite{Kane:2004,Visintin:2005} and the heat-induced formation of numerous gels such as biopolymer gels \cite{Nijenhuis:1997,FPTA:2006,Souguir:2015}, gels made of oil-in-water nanoemulsions \cite{Helgeson:2012,Helgeson:2014}, etc. In all of these examples, gelation is traditionally monitored in a geometry of constant gap with roughened boundary conditions to prevent wall slip. Small amplitude oscillatory shear (SAOS) is used to measure the linear elastic and viscous moduli, respectively $G'$ and $G''$, and the thermal dilation of the shear cell due to the temperature variation is taken into account through a gap compensation procedure that is now commonly implemented on rheometers.
 
Nevertheless, a key technical assumption in such measurements is that the contraction of the sample during the sol-gel (or the liquid-solid) transition is negligible. Such assumption is not always valid as phase transition often comes with a volume contraction of the sample. In the latter case, if the phase transition is monitored in a geometry of constant volume, the sample will experience strain hardening as it deforms and remains in contact with the cell boundaries. In a more subtle way, the contraction may also lead either to the complete debonding of the sample from the walls of the shear cell as observed for very rigid samples, or to the formation of lubrication patches at the gel/geometry interface for soft hydrogels \cite{Zhang:2001} which in both cases entail erroneous measurements of the sample elastic properties. Such issue is particularly important for biopolymer gels made of polysacharides and/or proteins which are the topic of the present work as they  experience volume variations during the sol-gel transition, and are prone to solvent release under external stress. 

 Quantitative measurements of the linear properties of such soft solids are crucial in numerous applications \cite{Gong:2010,Nguyen:2010}. For instance, biopolymer gels serve as substrate for bacterial growth and cell culture. The gel stiffness strongly impacts the cell locomotion, a phenomenon that is called ``durotaxis" \cite{Lo:2000}, as well as stems cell differentiation \cite{Engler:2006} which requires reliable methods to quantify the elastic properties of gels, especially when the latter is prepared through a complex thermal history. To our knowledge, most if not all the gelation experiments on thermoreversible gels reported in the literature have been performed at constant gap width, either in parallel-plate and cone-and-plate geometry \cite{Miyoshi:1996,Djabourov:1988,Mohammed:1998,Norziah:2006,Nordqvist:2011,Maurer:2012} or in a Taylor-Couette cell \cite{Richardson:1994,Normand:2000,Ikeda:2001}. None of these experiments takes into account the contraction of the sample during gelation, although the loss of adhesion between the gel and the wall of the shear cell has been already reported as a major issue \cite{Zhang:2001}. The use of sand paper, although widespread, do not compensate for the sample contraction. Moreover, the presence of such a medium sandwiched between the Peltier plate and the sample further impairs the temperature control of the sample, while the glue attaching the sandpaper poorly withstands high temperatures. As a result of the sample contraction, measurements of the viscoelastic moduli in different geometries are often inconsistent which urges experimentalists to use alternative methods to monitor more accurately non-isochoric processes such as the gelation of biopolymer gels.  

 The purpose of the present paper is twofold. First, it aims to illustrate in a series of benchmark experiments in the same spirit as a recent work by Ewoldt et al. \cite{Ewoldt:2015} that the use of a normal force controlled procedure is more adapted to monitor the gelation of thermoreversible gels than the traditional constant gap protocol. Normal force controlled rheology has been applied recently with success at constant temperature to the study of dense granular suspensions \cite{Mills:2009,Boyer:2011,Fall:2015}. However, if such method is also advertised by Anton Paar to probe soft solids in ``World of Rheology" \cite{Flow:2015}, or appears in the TA instruments help files, the present manuscript is to our knowledge among the first \textit{quantitative report} regarding normal force controlled rheology applied to thermoreversible gels. Here we show on a 1.5\% wt. agar gel, that applying a constant normal force equals to zero to the sample enclosed in a parallel-plate geometry prevents any strain hardening or debonding of the gel from the plates, thus providing precise measurements of both the gelation dynamics and the gel elastic properties. Furthermore, we show that such method works equally well with both rough and smooth boundary conditions, and that as such it can be used advantageously with bare metallic surfaces without any sandpaper.
Second, we build upon the first technical part and use the zero normal force protocol to study the impact of the thermal history of agar sols on the mechanical properties of the subsequent gels. We show that the cooling rate of the solution sets the gelation temperature and influences the gel contraction, without affecting significantly the final microstructure of the gel. Furthermore, the steady-state value of the gel elastic modulus is shown to be determined only by the final temperature reached at the end of the cooling ramp, and to be insensitive to the introduction of a constant temperature stage during the colling process. 

Finally, we highlight two artifacts that may take place during the gelation dynamics of agar gels, whether the gelation is monitored at constant gap or at constant normal force. First, we show that the slow oxidation of the metallic walls of the shear cell by the agar sol may lead to an artificial increase of $G'$ while the sample is still liquid. Second, we demonstrate that the oil rim surrounding the sample that is traditionally used to prevent the solvent evaporation, may slowly invade the gap also triggering a premature increase of the gel elastic modulus before the actual start of the gelation. Such an artifact leads to an erroneous estimate of the crossing of $G'$ and $G''$, but hardly impacts the measurement of the gel final elastic properties, as long as the terminal elastic modulus is larger than a few hundred pascals.

The paper is organized as follows. Section~\ref{Matmet} describes the agar samples, the experimental setup and the zero normal force procedure. We also discuss the effect of the strain amplitude and the impact of boundary conditions on the rheological measurements. In section~\ref{Results}, the zero normal force protocol is applied to determine the impact of thermal history on the mechanical properties of agar gels. Finally, section~\ref{Discussion} provides a short discussion of these results and emphasizes future applications of normal-force controlled rheology.

\begin{figure*}[!ht]
\includegraphics[width=\linewidth]{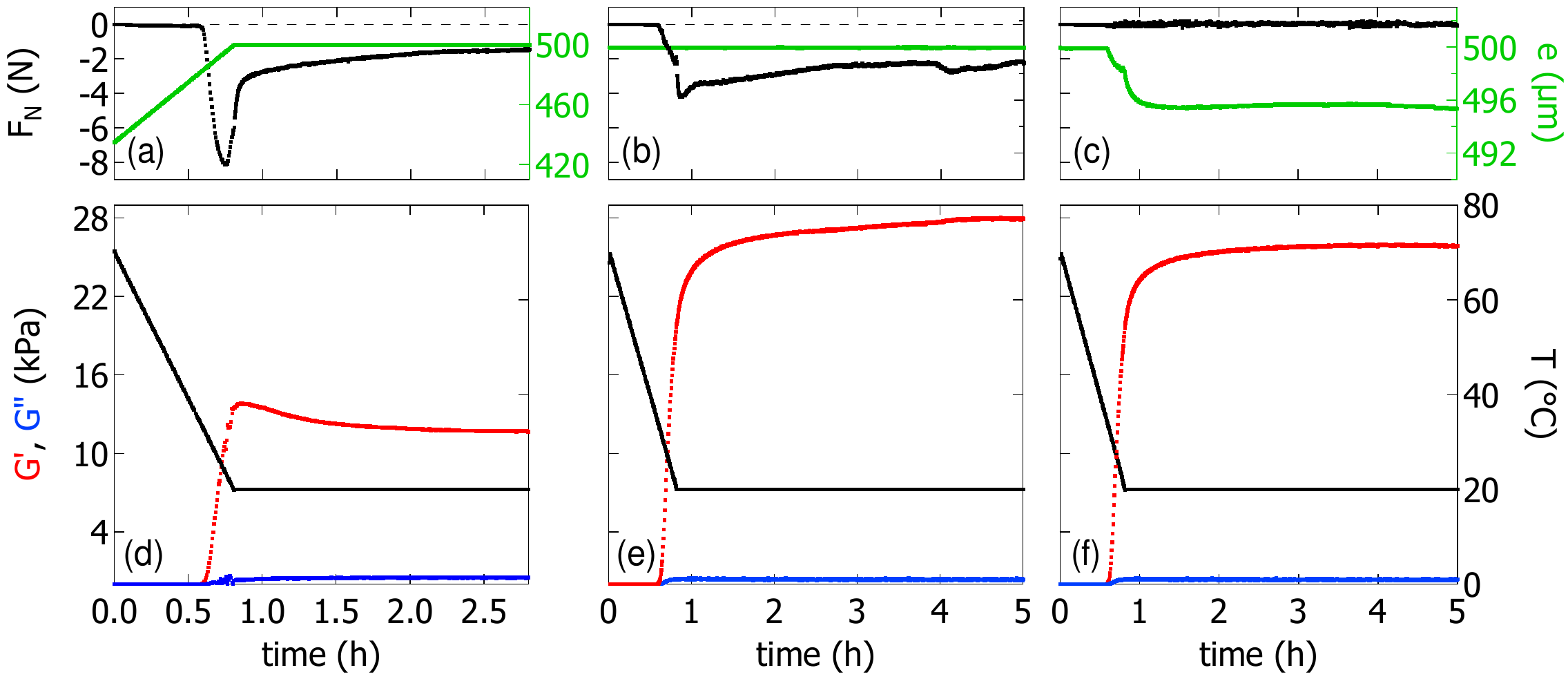}
\caption{\label{fig.1} (Color online) Temporal evolution of the normal force $F_N$ and the gap width $e$ (a)--(c) together with the elastic and viscous moduli (d)--(e) during the gelation of a 1.5\%~wt. agar solution induced by decreasing the temperature from $T=70^{\circ}$C to 20$^{\circ}$C at a cooling rate $\dot {\rm T}=1^{\circ}$C/min. The gelation experiment is repeated three times with different protocols. The first column illustrates the gelation performed without any particular precaution. Although the gap width is set initially at $e_0=500$~$\mu$m at $T=20^{\circ}$C, the true gap width varies by 65~$\mu$m due to the thermal dilation of the geometry, resulting in an  lower terminal value of the elastic modulus $G'_f=11.7$~kPa. The second column shows a gelation experiment during which the temperature compensation mode is active and compensates for the thermal dilation of the geometry alone. The gap width is truly constant during the whole gelation ($e_0=500$~$\mu$m). Yet, the normal force is still negative which proves that the sample contracts during the gelation. Such contraction leads to strain hardening as evidenced by the continuous increase of the elastic modulus ($G'_f\simeq 28.0$~kPa at $t=5$~h). Finally, the third column illustrates a gelation experiment during which a constant normal force is applied to the sample $F_N=(0.0\pm 0.1)$~N, while the temperature compensation mode is active. The gap width of initial value $e_0=500$~$\mu$m, decreases by 1\% which compensates for the sample contraction ($G'_f=25.9$~kPa). For the three experiments, the strain amplitude is $\gamma =0.01$~\%, the frequency is $f=1$~Hz and the solvent trap is filled with water.}%
\end{figure*}
    
\section{Material and methods}
\label{Matmet}

\subsection{Agar gels}

 Agar is extracted from a marine red algae and consists in a mixture of polysaccharides among which the gelling component is agarose, a polymer composed of two different galactose subunits \cite{Araki:1956,Stanley:2006}. When agar is dissolved in boiling water, the agarose is responsible for the formation of a crosslinked network upon cooling. The gelation involves a liquid-liquid phase separation known as spinodal decomposition that corresponds to the formation of polymer-rich and solvent-rich regions  \cite{Feke:1974,SanBiagio:1996,Matsuo:2002}, together with a conformational change of the agarose molecules which self-associate via hydrogen bonds \cite{Tako:1988,Braudo:1992}. The exact formation scenario of the gel results from a subtle competition between these three distinct processes \cite {Manno:1999} which in turn strongly depends on the agarose concentration, its molecular weight and the thermal history \cite{Normand:2000,Aymard:2001,Xiong:2005}. The spinodal demixion is favored for agarose concentrations lower than 2\% wt., whereas the gelation occurs through a more direct scenario at larger concentrations. Furthermore, although the exact conformation of the agarose molecules inside the gel network is still debated (single vs. double helices) \cite{Arnott:1974,Clark:1987,Foord:1989,Djabourov:1989,Schafer:1995,Guenet:2006}, the microstructure of agar gels at a coarser scale is consensually described as a porous network made of bundles of agarose chains \cite{Chui:1995,Pernodet:1997,Nordqvist:2011}. The latter microstructure is filled with water and displays a brittle-like mechanical behavior including the formation of macroscopic fractures at large enough strains \cite{Bonn:1998b,Barrangou:2006,Daniels:2007,Spandagos:2012}. Agar gels are also prone to release water at rest\cite{Matsuhashi:1990}, or under external deformation as low as a few pourcents \cite{Nakayama:1978} which makes their rheological study particularly challenging and motivates the present study.

 Here we focus on hydrogels made with 1.5~\%~wt of agar composed at 70\% of agarose (BioM\'erieux). Agar is added as a powder to warm deionized water, and the mixture is brought to a boil whilst stirring mildly for about 10 minutes. The solution is then cooled down to $T=80^{\circ}$C and kept at this temperature up to three days during which several samples are taken out to prepare gels inside the gap of a rheometer, as detailed below. After three days at $80^{\circ}$C, the polymer solution starts suffering chemical aging as evidenced by the formation of gels of lower elastic moduli \cite{Hickson:1968,Mao:2015}. Fresh solutions are therefore prepared regularly, and all the data reported here have been obtained from fresh batches, younger than three days.

\subsection{Rheological setup}
\label{MatmetRheo}
Rheological measurements are conducted in a plate-plate geometry of diameter 40~mm driven by a stress-controlled rheometer (DHR-2, TA Instruments). The upper moving plate made of duralumin is sand-blasted and displays a surface roughness of $4\pm 2$~$\mu$m, measured by Atomic Force Microscopy. The bottom plate consists in a smooth and Teflon-coated Peltier unit that allows us to control the temperature of the sample. Evaporation is minimized either by using a solvent trap containing deionized water and placed on top of the upper plate, or by adding a thin layer of sunflower seed oil from \textit{Helianthus annuus} (Sigma Aldrich) around the sample depending on the duration of the experiment (see section~\ref{evap} for more details). Gelation experiments are conducted as follows: the agar solution is first introduced at $T=80^{\circ}$C in the gap of the plate-plate geometry that has been preheated at $T=70^{\circ}$C. The temperature is then decreased at a constant rate $\dot {\rm T}=1^{\circ}$C/min down to $T_f=20^{\circ}$C (unless stated otherwise) and maintained at this temperature up to several hours to make sure the gelation is complete. In the meantime, oscillations of small amplitude performed at a frequency $f=1$~Hz allow us to monitor the temporal evolution of the gel viscoelastic properties. Gelation experiments are conducted either while imposing a constant gap width ($e=500~\mu$m), or a constant normal force ($F_N=0.0\pm0.1$~N) while the gap width is thus free to vary with the gel thickness. Finally, a few tests are performed on deionized water at constant temperature using either one or two transparent plates made of PMMA. In that case, images of the sample are taken from below with a webcam (Logitech HD c920) by means of a flat mirror placed at an angle of 45$^{\circ}$ with respect to the horizontal plates.

\subsection{Gap compensation and zero normal force protocol}

Let us start by illustrating the impact of the thermal expansion of the shear cell on the rheological measurements. For this first experiment, the gap width is set to $e=500~\mu$m at $T=20^{\circ}$C. However, due to the thermal dilation of the plates, the gap width decreases by 1.3~$\mu$m/$^{\circ}$C for increasing temperature, as determined by calibration. The gap width is therefore 435~$\mu$m at the moment the polymer solution is introduced inside the pre-heated geometry. Once the polymer solution is loaded, the temperature is decreased at a constant rate $\dot {\rm T}=1^{\circ}$C/min down to $T_f=20^{\circ}$C while we monitor the evolution of the normal force [Fig.~\ref{fig.1}(a)] and apply oscillations of small amplitude with a strain $\gamma =0.01$~\% to determine the elastic and viscous moduli of the gel, respectively $G'$ and $G''$ [Fig.~\ref{fig.1}(d)]. During the first half hour, while $T\geq 33^{\circ}$C the agar solution remains liquid, and the normal force $F_N$ is close to zero as the gap increases back to the initial value of 500~$\mu$m. At $t \simeq 0.6$~h, the gel starts forming as evidenced by the growth of the elastic modulus which becomes larger than the viscous modulus. Concomitantly, the normal force $F_N$ shows negative values indicating that the gel is pulling down on the upper plate, until a strong upturn occurs at $t=0.75$~h before the end of the cooling phase ($t=0.8$~h), indicating a brutal change in the contact between the gel and the plates, as discussed below. Moreover, $G'$ goes through a maximum at $t=0.88$~h after the temperature has reached $T_f=20^{\circ}$C, and relaxes towards a steady state value $G'_f =11.7$~kPa, while $F_N$ remains negative. The formation of the gel appears as over after 2.5~h, and the gel displays a solid-like behavior as evidenced by $G' \gg G''$.

In fact, the above procedure which does not compensate for the thermal dilation of the geometry leads to artificially low estimates of $G'$. Indeed, we have repeated the gelation experiment taking into account the variation of the gap thickness due to thermal dilation [Fig.~\ref{fig.1}(b)]. In that case, later referred to as the ``temperature compensation mode",  the gap width remains constant during the entire experiment and the terminal value of the elastic modulus is $G'_f=28.0$~kPa [Fig.~\ref{fig.1}(e)], instead of $11.7$~kPa in the absence of compensation for the thermal dilation of the geometry [Fig.~\ref{fig.1}(d)]. This experiment illustrates how dramatic and misleading can be the effect of the thermal dilation of the geometry on the measurements of the gel linear properties. As a consequence, the abrupt upturn of the normal force observed in Fig.~\ref{fig.1}(a) in the absence of any compensation results from a partial loss of contact between the sample and the geometry, and/or the formation of lubrication spots between the gel and the plate due to the large values of $|F_N|$. Indeed, from the maximum value of $|F_N| \sim 8$~N in Fig.~\ref{fig.1}(a), one can estimate the deformation experienced by the gel with an elastic modulus of 28~kPa to be of about 20\% which is large enough to trigger the release of solvent and the partial lubrication of the gel-plate interface. 

Nonetheless, the temperature compensation mode does not lead to fully satisfying rheological measurements. Indeed, one observes that the normal force still exhibits negative values during the gelation although we compensate for the thermal dilation of the geometry [Fig.~\ref{fig.1}(b)]. This result shows that the gel contracts during the sol-gel transition and pulls down on the upper plate. As a result, $G'$ displays a weak logarithmic increase even after $F_N$ has reached a plateau [Fig.~\ref{fig.1}(e)] which is the signature of the strain hardening of the agar gel induced by the sample contraction [see Fig.~\ref{fig.sup1bis} in the appendix for a supplemental experiment and the corresponding discussion]. Such a slow drift of $G'$ towards ever larger values is also visible in other studies conducted on gels containing agar(ose) \cite{Goycoolea:1995,Labropoulos:2002,Normand:2003,Piazza:2010}, but usually unnoticed and not discussed as rheological data are most often plotted in semilogarithmic scale which flattens $G'$ artificially.

\begin{figure*}[t]
\includegraphics[width=0.8\linewidth]{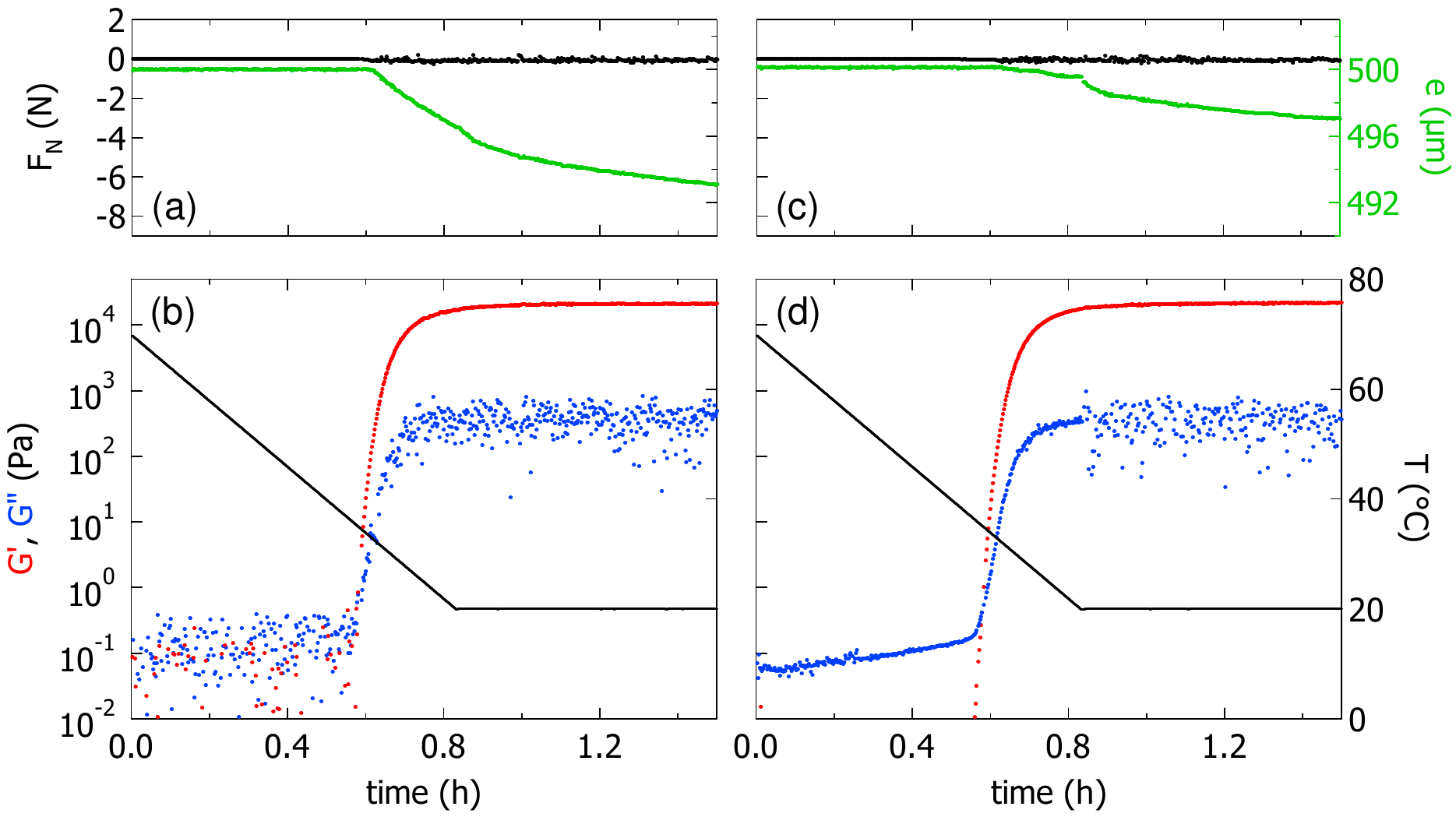}
\caption{\label{fig.2} (Color online) The first column shows the temporal evolution of (a) the gap width $e$ while imposing a constant normal $F_N=(0.0 \pm 0.1)$~N during the gelation of a 1.5\%~wt. agar solution induced by decreasing the temperature from $T=70^{\circ}$C to 20$^{\circ}$C at $\dot {\rm T}=1^{\circ}$C/min (relative decrease of the gap width $\Delta e/e = 1.4$\%). (b) Evolution of the elastic and viscous moduli in semilogarithmic scale determined through small amplitude oscillations with $\gamma =0.01$~\% and $f=1$~Hz (Crossover temperature of $G'$ and $G''$: $T_g=35.0^{\circ}$C; terminal value of the elastic modulus: $G'_f=21.0$~kPa). The second column shows the same experiment performed at $f=1$~Hz, but during which the strain $\gamma$ is adapted to the value of the elastic modulus for a better resolution from the early stage of the experiment while the sample is still liquid: the strain amplitude is $\gamma=1$~\% for $G'<1$~Pa, $\gamma=0.1$\% for 1~Pa $\leq G'<10$~Pa, and $\gamma=0.01$~\% for $G'\geq 10$~Pa ($\Delta e/e=0.6$~\%; $T_g=35.2^{\circ}$C; $G'_f=21.6$~kPa). For both experiments, the solvent trap is filled with water.
}%
\end{figure*}

 To compensate the effect of both the thermal expansion of the geometry and the contraction of the sample during the phase transition, the gel formation is now monitored with the temperature compensation mode switched on, while imposing a controlled normal force $F_N=(0.0\pm0.1)$~N instead of a constant gap width. The gap width, which initial value is set to $e_0=500$~$\mu$m may now vary during the experiment and indeed, we observe a gap decrease of 4~$\mu$m concomitantly to the growth of $G'$ [Fig.~\ref{fig.1}(c)]. Such a gap decrease of about 1\% allows the upper plate to stay in contact with the gel while the latter contracts during the gelation. Furthermore, the elastic modulus does not drift anymore and reaches a constant value $G'_f=25.9$~kPa at the end of the cooling phase [Fig.~\ref{fig.1}(f)], that is smaller than the one recorded in the constant gap experiment [Fig.~\ref{fig.1}(e)]. The latter observation also confirms the hardening scenario invoked in Fig.~\ref{fig.1}(e). The error bar associated with $G'_f$ is mainly given by the ability of the experimentalist to load the shear cell with the same amount of hot agar solution. Repeating the same experiment about ten times on samples extracted from different batches leads to compatible results within 10\%. There is thus only a slight difference between the value of $G'_f$ measured with a constant gap (after 5~h), and the value measured with a constant normal force \footnote{Note that repeating the gelation experiment under zero normal force with other initial gap values, i.e. $e_0=200$~$\mu$m and 1000~$\mu$m gives compatible values of $G'_f$ within error bars, which demonstrates that the value of $G'_f$ determined with the zero normal force protocol is independent of the initial value $e_0$ of the gap width.}. However the value of $G'_f$ measured with a constant gap width protocol is a function of time and as such depends on the patience of the experimentalist. In conclusion, monitoring the gelation while imposing a constant normal force equals to zero prevents the strain hardening of the gel and appears as a more accurate method to determine the gel elastic properties. This method, that is further referred to as ``the zero normal force protocol" is the one that is used in the rest of the manuscript. 

In the rest of section~\ref{Matmet} we show that the terminal value of $G'$ determined with the zero normal force protocol is robust and independent of the surface roughness of the plates. We also discuss in depth several artifacts related to the strain amplitude applied to measure the viscoelastic moduli, and to physico-chemical phenomena related to the boundary conditions. Finally, in section~\ref{Results} we use the zero normal force protocol to investigate the impact of thermal history on the linear properties of a 1.5\% wt. agar gel.

\begin{figure*}[t]
\includegraphics[width=0.8\linewidth]{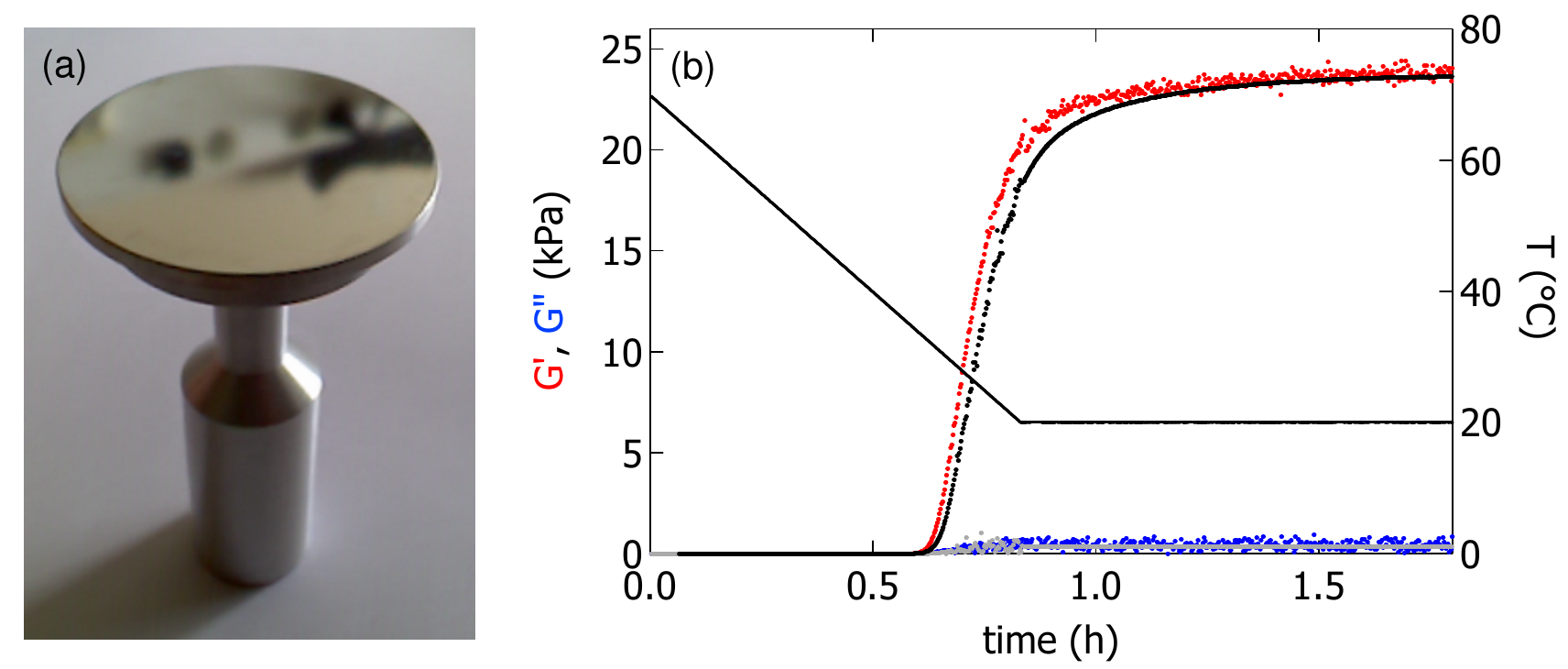}
\caption{\label{fig.2c} (Color online) (a) Picture of the duralumin rotor which has been polished with a suspension of ultrafine aluminium oxide particles. The scale is fixed by the plate diameter which is 40~mm. (b) Evolution of the viscoelastic moduli $G'$ and $G''$ vs. time during the gelation of a 1.5\%~wt. agar solution induced by decreasing the temperature from $T=70^{\circ}$C to 20$^{\circ}$C at $\dot {\rm T}=1^{\circ}$C/min, under controlled normal force $F_N=(0.0\pm0.1)$~N (initial gap size $e_0=500$~$\mu$m). The experiment is performed twice, each time with symmetric boundary conditions: once with smooth walls, i.e. a polished duralumin upper plate pictured in (a) and a Teflon coated bottom plate (red and blue symbols) and once with rough walls, i.e. sand-blasted duralumin plates (black and gray symbols). In both cases the relative gap decrease is 1\%, the crossover temperature of $G'$ and $G''$ is $T_g=35.7^{\circ}$C, and the terminal values of the elastic modulus are respectively $G'_f=23.8$~kPa and $23.6$~kPa. Both experiments are perfomed with a solvent trap filled with water.  
}%
\end{figure*}

\subsection{Strain adapted protocol}
\label{BC}

Here we briefly discuss the impact of the strain amplitude $\gamma$ applied to monitor the gel formation. Fig.~\ref{fig.2}(b) displays in a semi-logarithmic plot the temporal evolution of the elastic and viscous moduli during a gelation conducted with the zero normal force protocol and determined with $\gamma=0.01$~\% and $f=1$~Hz [same protocol as in Fig.~\ref{fig.1}(f)]. Although the sample is liquid-like during the first half hour, such a small deformation leads to noisy and comparable values of $G'$ and $G''$ until both moduli exceed about 1~Pa. This experimental observation, which is commonly encountered in the literature \cite{Altmann:2004,Russ:2014}, is a consequence of the too small applied value of the strain amplitude. Indeed, the minimum measurable value of $G'$ is inversely proportional to $\gamma$ and given by the following expression: $G'_{\rm min}= 4 T_{\rm min}/(3\pi R^3 \gamma)$, where $R$ denotes the plate radius and $T_{\rm \min}$ the smallest measurable torque \cite{Ewoldt:2015}. Here, $T_{\rm min}=2$~nN.m (for a DHR-2) and with $\gamma=0.01$~\% we find $G'_{\rm min}=1$~Pa which is in excellent agreement with the minimal significant stress value observed in Fig.~\ref{fig.2}(b).
Imposing a larger strain before the start of the gelation, such as $\gamma=1$~\% allows us to decrease $G'_{\rm min}$ down to 0.01~Pa and measure consistent values for the viscoelastic moduli during the first half hour, i.e. $G'=0$ since the sample is liquid [Fig.~\ref{fig.2}(d)]. The strain amplitude is then decreased when $G'$ increases as the gelation starts, so as to remain in the linear regime and prevent any strain-induced release of solvent and/or debonding of the sample from the geometry. Indeed, strain sweep experiments performed on gels which have been prepared under oscillations of different amplitudes $\gamma$ indicate that strain amplitudes as low as $\gamma=$0.01\% are necessary for an optimal adhesion between the gel and the plates (see Fig.~\ref{fig.sup2} and related discussion in the appendix which proves that the gel adhesion to the plates decreases for increasing strain values $\gamma$ applied during the gelation). As a consequence, the following protocol leads to reproducible measurements: $\gamma=1$~\% for $G'<1$~Pa, $\gamma=0.1$~\% for $1$~Pa$<G'<10$~Pa and $\gamma=0.01$~\% for $G'>10$~Pa, and confirms that as long as the solution is liquid, only $G''$ displays non-negligible values. After gelation, the terminal value of $G'$ obtained with the strain adapted protocol is compatible within error bars with the value measured with a constant strain $\gamma=0.01$~\% [Fig.~\ref{fig.1}(f)]. Therefore adapting the value of strain amplitude during the gelation offers the opportunity to determine more accurately the crossover between $G'$ and $G''$ which is often used as a good estimate of the gelation point \footnote{Note that a proper definition of the gelation point is the instant where $G'$ and $G''$ both scale as identical power laws of frequency which corresponds to the value of the phase angle $\delta = \arctan(G''/G')$ that is independent of the frequency \cite{Chambon:1987}. However, our goal here is not to determine a gelation point, but only to improve the determination of the intersection of $G'$ and $G''$.} and which occurs here at $T_g=35.2^{\circ}$~C. Such a strain-adapted protocol is systematically applied during the experiments reported in the rest of the manuscript.

\subsection{Boundary conditions and related artifacts}

\subsubsection{Surface roughness} \label{roughness}
Let us now discuss the impact of the surface roughness of the plates. The experiments reported in Fig.~\ref{fig.1} and~\ref{fig.2} have been performed with mixed boundary conditions (BC), namely a rough metallic upper plate and a smooth (Teflon-coated) bottom Peltier plate. To further test the robustness of the zero normal force protocol, we have performed gelation experiments with symmetric boundary conditions. On the one hand, experiments with symmetrically smooth boundary conditions are performed by replacing the sand-blasted duralumin rotor with a duralumin plate polished on a rotating wheel with a suspension of ultrafine aluminum oxide particles to produce a mirror like surface pictured in Fig.~\ref{fig.2c}(a) and of typical roughness $10\pm15$~nm, as determined from AFM measurements. The bottom plate is the smooth and Teflon-coated Peltier plate used by default. On the other hand, experiments with symmetrically rough boundary conditions are performed by topping the Peltier plate with a sandblasted duralumin cover that ensures a good thermal conductivity between the Peltier plate and the sample, and allows us to achieve a roughness of about 4~$\mu$m for the bottom plate similar to that of the upper sandblasted duralumin plate. Both experiments reported in Fig.~\ref{fig.2c}(b) lead to compatible values of $G'_f$ within error bars (rough BC: $G'_f=23.6$~kPa; smooth BC: $G'_f=23.8$~kPa) that are also in quantitative agreement with the elastic modulus determined with mixed boundary conditions [see Fig.~\ref{fig.1}(f)]. These experiments demonstrate that the zero normal force protocol together with oscillations of small amplitude adapted to the value of the elastic modulus allows one to monitor the gelation dynamics in a plate-plate geometry with either smooth or sand-blasted boundary conditions. The zero normal force protocol thus appears as much simpler than the traditional constant gap width protocol which requires the use of sandpaper. In particular, sandpaper does not compensate the sample contraction during gelation and is a poor thermal conductor. Moreover, it often displays adhesion issues to the Peltier plate at high temperature, hence inducing large error bars.  

\begin{figure*}[t]
\includegraphics[width=0.8\linewidth]{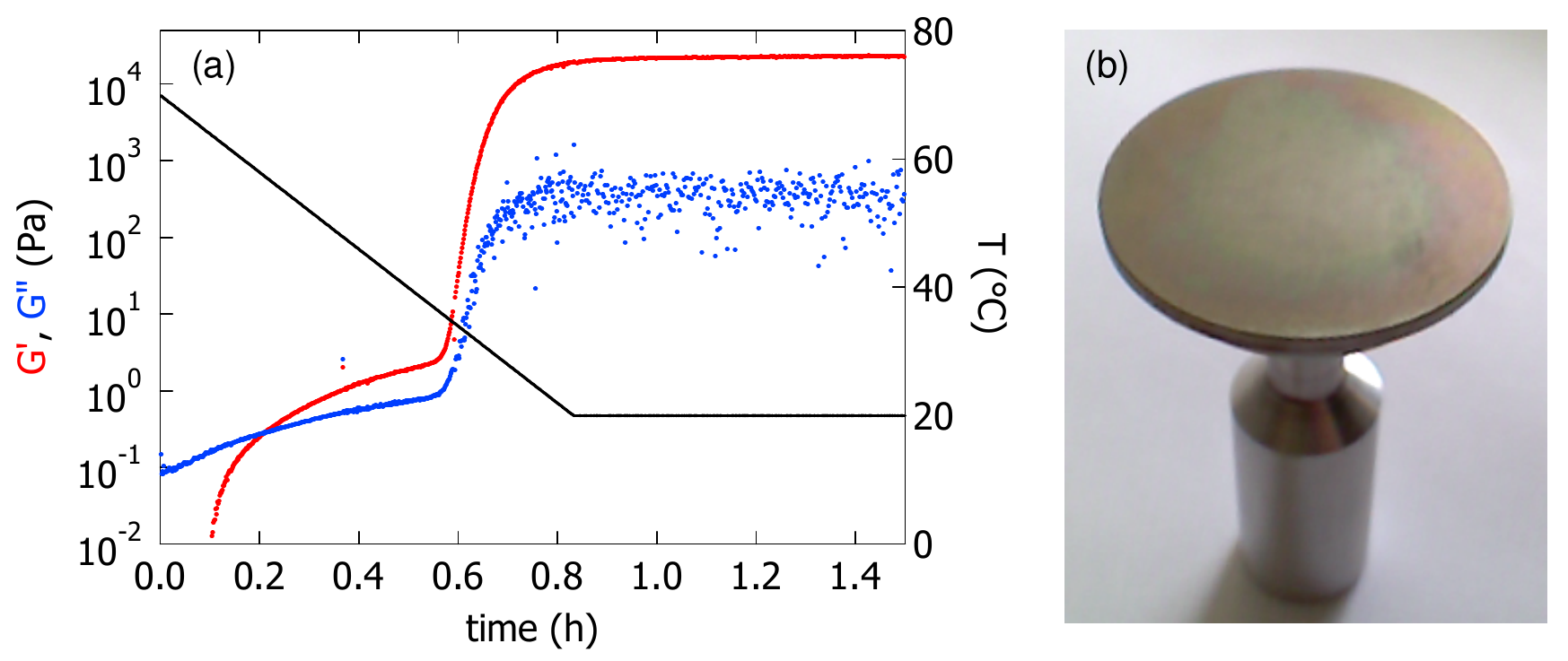}
\caption{\label{fig.2b} (Color online) (a) Temporal evolution of $G'$ and $G''$ during the gelation of a 1.5\%~wt. agar solution induced by decreasing the temperature from $T=70^{\circ}$C to 20$^{\circ}$C at a cooling rate $\dot {\rm T}=1^{\circ}$C/min and under controlled normal force $F_N=(0.0\pm0.1)$~N (initial gap size $e_0=500$~$\mu$m, relative decrease of the gap width: $\Delta e/e = 1$\%); terminal value of the elastic modulus: $G'_f=23.1$~kPa. The duralumin rotor has been sandblasted just before the experiment, and the gelation is performed with a solvent trap filled with water. (b) Picture of the sandblasted rotor after a few successive gelation experiments. Notice the iridescent regions at the periphery of the plate which corresponds to the oxidized area. The scale is set by the rotor diameter of 40~mm.}%
\end{figure*}

\subsubsection{Chemical properties of the plates surfaces} \label{Chemy}
 This subsection concerns a subtle artifact associated with the use of duralumin plates. We have observed that gelations performed with a duralumin rotor which has been freshly sandblasted before the experiment, systematically show a premature increase of the elastic modulus. An example is pictured in Fig.~\ref{fig.2b}(a): $G'$ increases from $t\simeq 0.1$~h instead of $t=0.55$~h [compare with Fig.~\ref{fig.2}(d)], and tends toward about 2~Pa before following the steep increase previously associated with the gelation. Such a premature increase of $G'$ progressively disappears when the experiment is repeated four or five times with the very same geometry, while the upper plate takes an iridescent appearance [Fig.~\ref{fig.2b}(b)]. If the rotor is sandblasted anew, the premature increase of $G'$ is again visible in the subsequent first couple of experiments. We attribute such a premature increase of $G'$ to the oxidation of the upper duralumin plate by the agar aqueous solution. Nanoparticles of aluminium oxide detach from the surface of the upper plate to be advected through the sample toward the air/liquid interface, which increases the elasticity of the contact line and leads to the premature increase of $G'$. In this framework, the successive gelation experiments result in the passivation of the upper plate which strongly slows down any further oxidation of the geometry and explains why the premature increase of $G'$ is no longer observed in the subsequent experiments.

\begin{figure*}[t]
\includegraphics[width=0.7\linewidth]{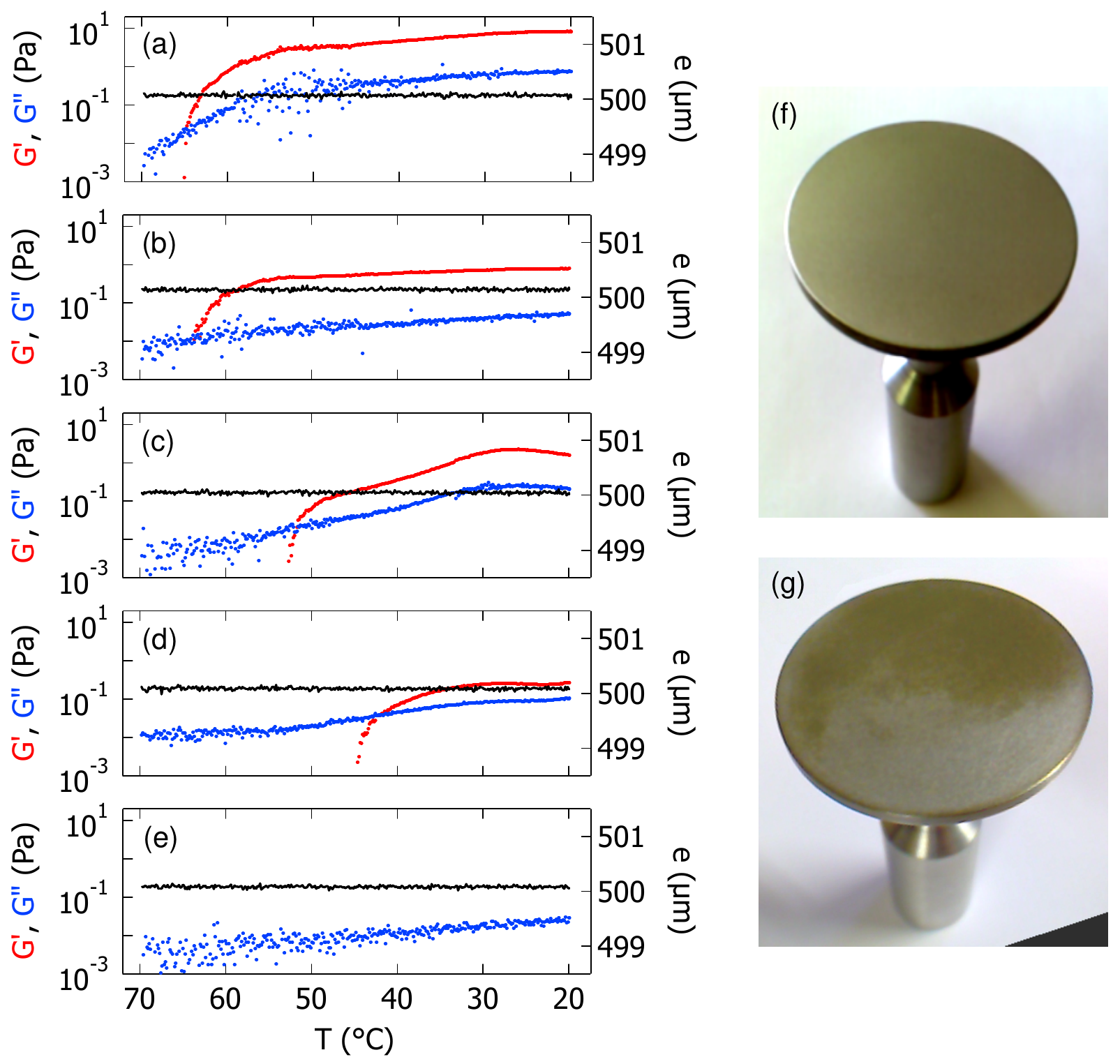}
\caption{\label{fig.2d} (Color online) (a)-(e) Temporal evolution of $G'$ and $G''$ of distilled water during a series of five consecutive experiments with mixed boundary conditions (rough duralumin upper plate and smooth bottom Peltier plate). Each experiment corresponds to a new water sample and consists in oscillations of small amplitude ($\gamma=1$~\%, $f=1$~Hz) during a decreasing ramp of temperature from $T=70^{\circ}$C to 20$^{\circ}$C at a cooling rate $\dot {\rm T}=1^{\circ}$C/min with a solvent trap is filled with water. The zero normal force protocol is applied with $F_N=(0.0\pm 0.1)$~N and an initial gap value  $e_0=500$~$\mu$m. Note that during each experiment, the gap remains constant equal to the initial value, which proves that there is no evaporation. The rotor has been sandblasted anew only before the first experiment displayed in (a). Pictures of the  duralumin rotor: (f) before the first experiment and (g) after the fifth experiment. The scale is set by the rotor diameter of 40~mm.}%
\end{figure*}

 To support this interpretation, we have performed the following test: the rotor is sandblasted anew, and the preheated gap is filled with distilled water instead of the agar sol. The temperature is decreased as usual from $T=70^{\circ}$C down to 20$^{\circ}$C at a cooling rate $\dot {\rm T}=1^{\circ}$C/min under a constant normal force $F_N=$($0.0 \pm 0.1$)~N and oscillations of small amplitude ($\gamma=1$~\% and $f=1$~Hz). Although no gelation can take place, one can see in Fig.~\ref{fig.2d}(a) that after 5~min, the elastic modulus increases up to a few Pa, and remains larger than $G''$ until the end of the temperature ramp. The same experiment is repeated four more times with fresh distilled water, and the results are pictured in Fig.~\ref{fig.2d}(b)-(e). The increase of $G'$ is still visible, but delayed from one experiment to the next and eventually no longer observed during the fifth experiment [Fig.~\ref{fig.2d}(e)]. Furthermore, comparing the surface state of the upper plate before the first experiment [Fig.~\ref{fig.2d}(f)] and after the fifth experiment [Fig.~\ref{fig.2d}(g)], one observes an oxide layer visible by the naked eye in the latter case. The plate is indeed covered by some powdery material that can be removed easily with the finger which confirms the scenario discussed above: the corrosion of the duralumin plate nucleate aluminum oxide particles that increase the contact line elasticity leading to the premature increase of $G'$. Note that the bottom Peltier plate used in all the experiments reported in the present article (except in section~\ref{roughness}) is Teflon-coated which explains why the oxidation is only taking place at the rotor. 

In conclusion, one should be aware that sandblasting a duralumin plate makes it rough, but also promotes its slow oxidation by the agar sol, which leads to an artificial increase of $G'$ that should not be interpreted as the early onset of the gel formation. We shall emphasize that such issue is absent when using commercial plates made of stainless steal. In section~\ref{Results}, all the gelation experiments are performed with an upper plate made of passivated duralumin to prevent any premature increase of $G'$.

\subsection{Solvent evaporation, oil coating and related artifacts}
\label{evap}

\begin{figure*}
\includegraphics[width=0.8\linewidth]{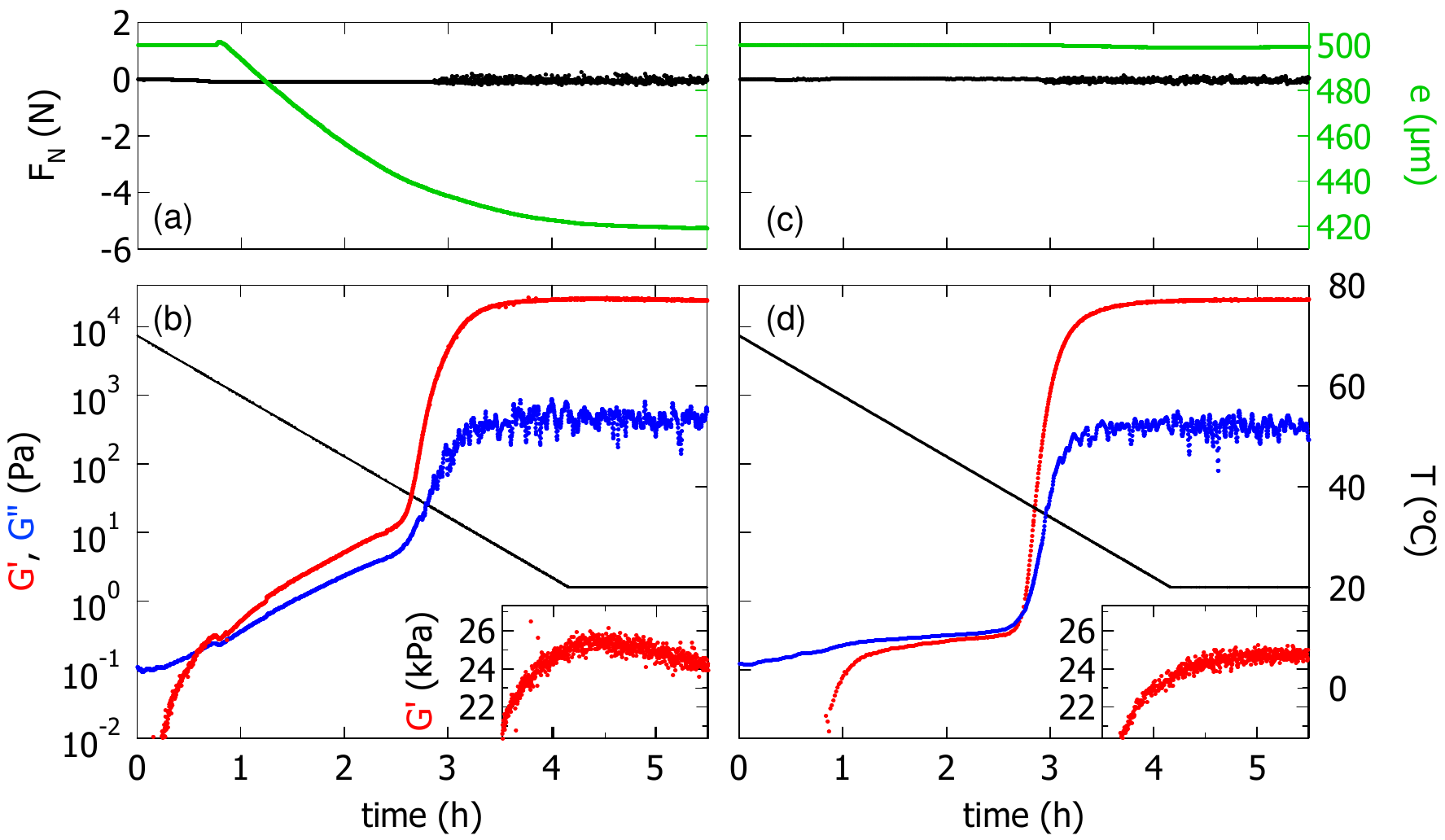}
\caption{\label{fig.3} (Color online) Gelation dynamics of a 1.5\%~wt. agar solution induced by decreasing the temperature from $T=70^{\circ}$C to 20$^{\circ}$C at a low cooling rate $\dot {\rm T}=0.2^{\circ}$C/min under a controlled normal force $F_N=(0.0\pm0.1)$~N. Temporal evolution of (a) the gap size $e$ (initial value $e_0=500~\mu$m) and (b) the viscoelastic moduli $G'$ and $G''$ in semilogarithmic scale measured at $f=1$~Hz with the adapted strain protocol and an oscillatory frequency of $f=1$~Hz. The inset is a zoom of $G'$ vs time plotted in linear scale for $3.5<t<4.5$~h. The solvent trap is filled with water. (c) and (d) same as in (a) and (b) except that the sample is surrounded by a thin oil layer. 
}%
\end{figure*}

Agar gels are mainly composed of water and evaporation has a strong impact on both the gelation dynamics and the gel linear properties. For gelation experiments performed under rapid cooling rates, i.e. $\dot {\rm T}\geq 1^{\circ}$C/min, evaporation is efficiently prevented by a solvent trap filled with water. However, for long lasting experiments conducted at lower cooling rates, evaporation becomes non negligible despite the solvent trap. A trick commonly reported in the literature consists in adding a thin layer of a non volatile oil at the periphery of the plate to isolate the sample from the ambiant air \cite{Orafidiya:1989,Mohammed:1998,Nordqvist:2011}. The purpose of this section is to demonstrate that such method efficiently prevents the evaporation and to investigate quantitatively the impact of the oil layer on the rheological measurements. 

To this end, we perform two gelation experiments of long duration at a cooling rate $\dot {\rm T}=0.2^{\circ}$C/min using a passivated sand-blasted duralumin rotor and the zero normal force protocol. The first gelation experiment is performed with water in the solvent trap [Fig.~\ref{fig.3}(a) and (b)], while in the second gelation experiment the sample is surrounded by a thin oil layer [Fig.~\ref{fig.3}(c) and (d)]. In the first case, we observe visually that the water in the solvent trap topping the rotor has evaporated after about 20~min, which coincides with the time above which $G'$ shows non-negligible values [Fig.~\ref{fig.3}(b)]. The elastic and viscous moduli intersect at $t\simeq0.6$~h while the gap starts decreasing. Both $G'$ and $G''$ slowly grow for about 2~h before showing a steep increase associated with the gelation. Non-negligible evaporation is further supported by two other observations. First, the large gap decrease of about 16\% illustrated in Fig.~\ref{fig.3}(a) shows that the sample experiences water loss and that the increase of $G'$ and $G''$ between 0.5 and 2.5~h is artificial. Indeed, in the second experiment where the evaporation remains negligible because of the oil surrounding the sample, the gap decreases by less than 1\% over the entire experiment [Fig.~\ref{fig.3}(d)]. 
Second, in the experiment conducted with water in the solvent trap, $G'$ goes through a maximum at $t\simeq4.2$~h before decreasing at larger times [Fig.~\ref{fig.3}(b)--inset], whereas in the experiment using oil to prevent evaporation, $G'$ exhibits a monotonic behavior and tends toward a constant value after 5~hours [Fig.~\ref{fig.3}(d)--inset].

These results demonstrate quantitatively that for a low cooling rate of $0.2^{\circ}$C/min, the sample experiences excessive evaporation when the solvent trap is filled with water. The use of an oil layer at the sample periphery prevents the solvent evaporation and allows one to access the steady state elastic properties of the gel \footnote{We checked that the surrounding oil does not impact the steady state value of the elastic modulus $G'_f$ by performing an experiment over a much shorter duration. Two experiments conducted at a cooling rate $\dot {\rm T}=1^{\circ}$C/min, one with water in the solvent trap and the other one with an oil layer around the sample lead to compatible values of the elastic modulus, within error bars.}. However, if the use of oil leads to quantitative measurements in steady state, $G'$ shows non negligible values from $t\simeq 0.85$~h while the sample is still liquid, and long before the steep increase attributed to the gelation which takes place at $t\simeq 2.7$~h [Fig.~\ref{fig.3}(d)]. Such behavior is reminiscent of the artifact described in the previous subsection~\ref{Chemy}. Nonetheless, the experiments reported in Fig.~\ref{fig.3} are conducted with a passivated duralumin rotor, thus the premature increase of $G'$ is here related to another mechanism than the corrosion of the upper plate. We show in the remainder of this section that the premature growth of $G'$ is an artifact related to the presence of the surrounding oil layer.

\begin{figure*}[t]
\includegraphics[width=\linewidth]{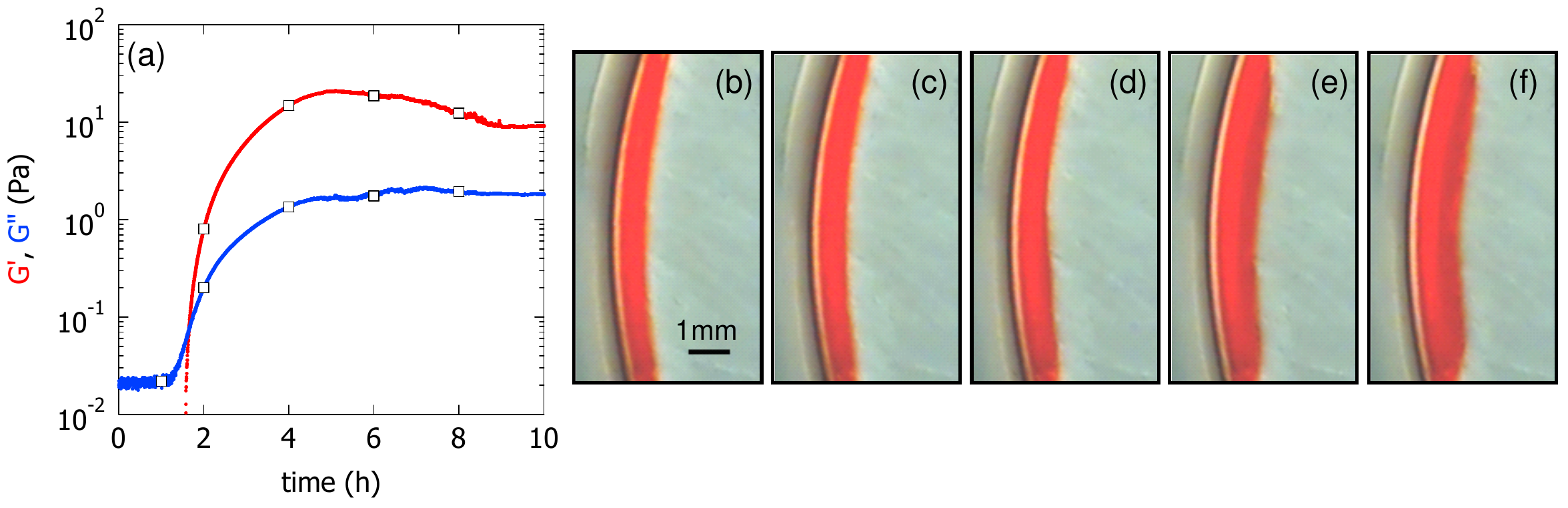}
\caption{\label{fig.4} (Color online) (a) Evolution of the elastic and viscous moduli of deionized water surrounded by a thin layer of sunflower seed oil. Measurements are performed at $T=20^{\circ}$C in a plate-plate geometry with a constant gap $e=500~\mu$m, under small amplitude oscillatory shear ($\gamma=1$~\% and $f=1$~Hz). The bottom plate is a transparent piece of PMMA, while the upper plate is made of sandblasted and passivated duralumin. (b)-(f) Snapshots of the periphery of the plate taken from below, through the transparent bottom plate at times $t=1$, 2, 4, 6 and 8~h [these times are indicated by open symbols ($\square$) in graph (a)]. The images show that the surrounding oil which is stained in red with a fat soluble dye (Sudan III), slowly invades the gap.}%
\end{figure*}

To study the effect of the surrounding oil alone, i.e. in the absence of the gelation, we perform a time sweep experiment on deionized water surrounded by a thin layer of oil. The experiment is conducted at constant temperature ($T=20^{\circ}$C) using a passivated duralumin upper plate and a transparent bottom plate made of PMMA (gap $e=500$~$\mu$m). Although no gelation can take place, $G'$ increases and becomes larger than $G''$ from $t>1.6$~h to reach a value $G' \simeq 10$~Pa after six hours [Fig.~\ref{fig.4}(a)]. Moreover, images taken through the transparent bottom plate shows that concomitantly to the growth of $G'$, the oil rim slowly invades the gap at certain location along the sample periphery, and spreads on the surface of the upper metallic plate [Fig.~\ref{fig.4}(b)-(f)]. Therefore, the premature growth of $G'$ reported in Fig.~\ref{fig.3}(d) during the gelation of the agar sol is most likley related to the inner migration of the oil/water interface.  

\begin{figure*}[t]
\includegraphics[width=\linewidth]{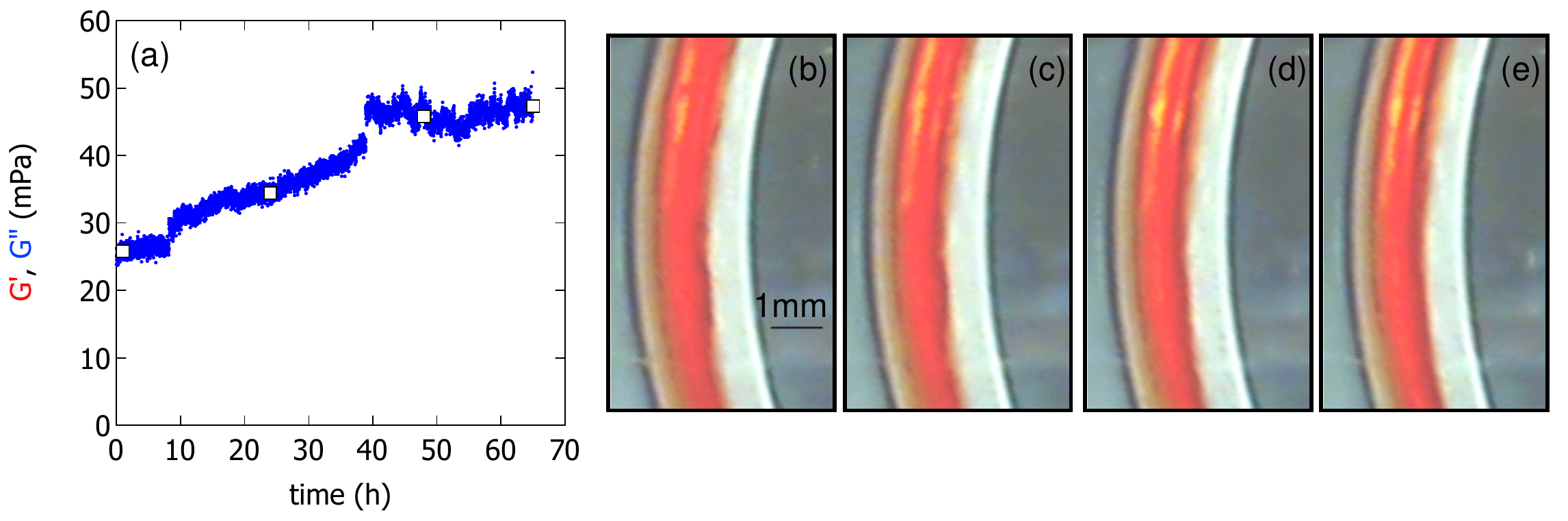}
\caption{\label{fig.4b} (Color online) (a) Temporal evolution of the elastic and viscous moduli of deionized water surrounded by a thin layer of sunflower seed oil. Measurements are performed at $T=20^{\circ}$C in a plate-plate geometry with a constant gap $e=500~\mu$m, under small amplitude oscillatory shear ($\gamma=1$~\% and $f=1$~Hz). The bottom and the upper plate are made of PMMA and transparent. (b)-(e) Snapshots of the periphery of the plate taken from below, through the transparent bottom plate at times $t=1$, 24, 48 and 65~h [these times are indicated by open symbols ($\square$) in graph (a)]. The surrounding oil is stained in red with a fat soluble dye (Sudan III).
}%
\end{figure*}

To further explore the role of the upper metallic plate on such an invasion process, we repeat the time sweep experiment on deionized water surrounded by a thin oil layer, with symmetric boundary conditions using two identical transparent plates made of PMMA (constant temperature: $T=20^{\circ}$C and constant gap: $e=500~\mu$m). For more than 65~h, $G'=0$ and only $G''$ shows non-negligible values [Fig.~\ref{fig.4b}(a)], while no migration of the oil/water interface is observed by the naked eye [Fig.~\ref{fig.4b}(b)-(e)]. These results confirm that the growth of $G'$ and the migration of the oil/water interface observed in Fig.~\ref{fig.4} are correlated. Furthermore it shows that the invasion of the gap by the outer oil layer, and therefore the premature increase of $G'$, are directly related to the use of metallic boundary conditions. 

Finally, if a more detailed analysis of the mechanism responsible for the oil invasion of the gap is out of the scope of the present publication, we discuss the weak linear increase of $G''$ visible in Fig.~\ref{fig.4b}(a) which is instructive. The increase of $G''$ results from the slow pervaporation of water through the oil layer. Indeed, a careful spatio-temporal analysis of the position of the oil/water interface pictured in Fig.~\ref{fig.4b}(b)-(e) shows a net displacement $\delta l$ of the oil/water interface after 65 hours \footnote{Note that one can also estimate the length $\delta l$ over which the oil invades the gap from the increase of the loss modulus $G''$. Indeed, since $G''$ corresponds to the energy dissipated per unit volume, at first order the relative variation of $G''$ follows:
\[\frac{G''(\delta l)-G''(\delta l=0)}{G''(\delta l=0)} = 4\left(\frac{\eta_o}{\eta_w}-1 \right)\frac{\delta l}{R} \simeq 4\frac{\eta_o}{\eta_w}\frac{\delta l}{R},\] \\ with \[ G''\simeq \int_0^R \eta \dot \gamma^22\pi r e dr, \] where $\dot \gamma=\Omega r/e$ is the local shear rate, $\eta$ is the viscosity of the liquid phase (with $\eta=\eta_w$ for $r<R-\delta l$ and $\eta=\eta_o$ for $r> R-\delta l$, where $\eta_o$ and $\eta_w$ stand respectively for the viscosity of the surrounding oil and that of water), $R$ denotes the plate radius and $\Omega$ is the instantaneous angular velocity of the upper plate. From Fig.~\ref{fig.4b}(a), one infers that $\Delta G''/G''(\delta l=0) \simeq 0.7$ and with $\eta_o=60\eta_w$ one finds $\delta l\simeq 50\mu$m in excellent agreement with the average radial displacement of the oil/water interface derived from the spatiotemporal analysis of the snapshots pictured in Fig.~\ref{fig.4b}(b)-(e) (see Fig.~\ref{fig.sup1} and discussion in the appendix).}
ranging between 30 and 50~$\mu$m depending on the angular position along the interface [see Fig.~\ref{fig.sup1} in the appendix]. Such a displacement corresponds to an average velocity of the oil/water interface of about 0.17~nm.s$^{-1}$. Assuming a near-diffusive transfer, the migration velocity $v$ of the oil/water interface should scale as $P D/\xi$, where $D$ is the diffusion coefficient of water in oil, $P$ is the partition coefficient of water at the oil/water interface and $\xi$ denotes the thickness of the oil layer \cite{Ziane:2015}. For water permeation through a vegetable oil/water interface \footnote{Note that the diffusion coefficient $D\simeq 10^{-10}$~m$^2$.s$^{-1}$ of water in sunflower seed oil is an order of magnitude lower than the diffusivity of water in silicone oil reported in \cite{Ziane:2015} since the polar moieties of triglycerides strongly interact with water through hydrogen bonds, resulting in a lower mobility of dissolved water molecules \cite{Zieverink:2009}.} one has $P\simeq 1.6\times 10^{-3}$ and $D\simeq 10^{-10}$~m$^2$.s$^{-1}$ \cite{Zieverink:2009}. Considering an oil layer of thickness $\xi\simeq 1$~mm [Fig.~\ref{fig.4b}(b)], the above relation leads to $v \simeq 0.16$~nm.s$^{-1}$ in very good agreement with the experimental observations, which confirms that the pervaporation scenario accounts for the increase of $G''$ in Fig.~\ref{fig.4b}(a).

In conclusion, using a layer of oil to surround the sample efficiently prevents the evaporation and allows one to measure reliable steady-state values of the viscoelastic moduli. Nonetheless, one should keep in mind that the presence of an oil layer together with metallic boundary conditions may result in the slow migration of the oil/water interface at the periphery of the plates, in turn leading to the premature growth of $G'$. As a consequence, any early evolution of the elastic modulus during the formation of an agar(ose) gel monitored with a surrounding oil layer should be considered with caution. In particular, the premature growth of $G'$ reported in ref. \cite{Nordqvist:2011} in the early stage of a gelation experiment is likely to be artificial and may not be the signature of the formation of an intermediate structure and/or a pre-gel inside the gap.


\section{Impact of thermal history on the gel mechanical properties}
\label{Results}

In section~\ref{Matmet}, we have shown that the zero normal force protocol is an efficient way to monitor the heat-induced formation of agar gels. This method, which may be used with either rough or smooth boundary conditions, prevents the loss of contact between the gel and the plates and allows one to measure reliably the terminal values of the gel elastic and viscous moduli. In what follows, the zero normal force protocol is applied to determine the influence of the thermal history on the linear properties of a 1.5\% wt. agar gel. We first quantify the impact of the cooling rate $\dot{\rm{T}}$ on the gel elastic properties, before turning to the role of the final temperature T$_f$ reached at the end of the linear ramp, for a fixed cooling rate of 1$^{\circ}$C/min. These experiments are conducted with the mixed boundary conditions described in section~\ref{MatmetRheo}: the upper duralumin plate is rough and passivated to prevent any oxidation (see subsection~\ref{Chemy}), while the bottom Peltier plate is smooth and Teflon coated. Unless stated otherwise, the gelation experiments reported in this last section are conducted using an oil layer around the sample to prevent evaporation (see section~\ref{evap}).

\subsection{Impact of the cooling rate}
\label{coolingrate}

In order to determine the impact of the cooling rate $\dot {\rm T}$ on the gel elastic properties, we have performed a series of gelation experiments on a 1.5\%~wt agar solution, at different cooling rates ranging from 0.1 to $10^{\circ}$C/min. For each experiment, the temperature is decreased from $70^{\circ}$C down to $T_f=20^{\circ}$C. The zero normal force protocol is applied to monitor the gelation, and the value of the strain is further adapted to the value of the elastic modulus as discussed in~\ref{BC}, to properly determine the intersection of $G'$ and $G''$. 

\begin{figure*}[t!]
\includegraphics[width=0.9\linewidth]{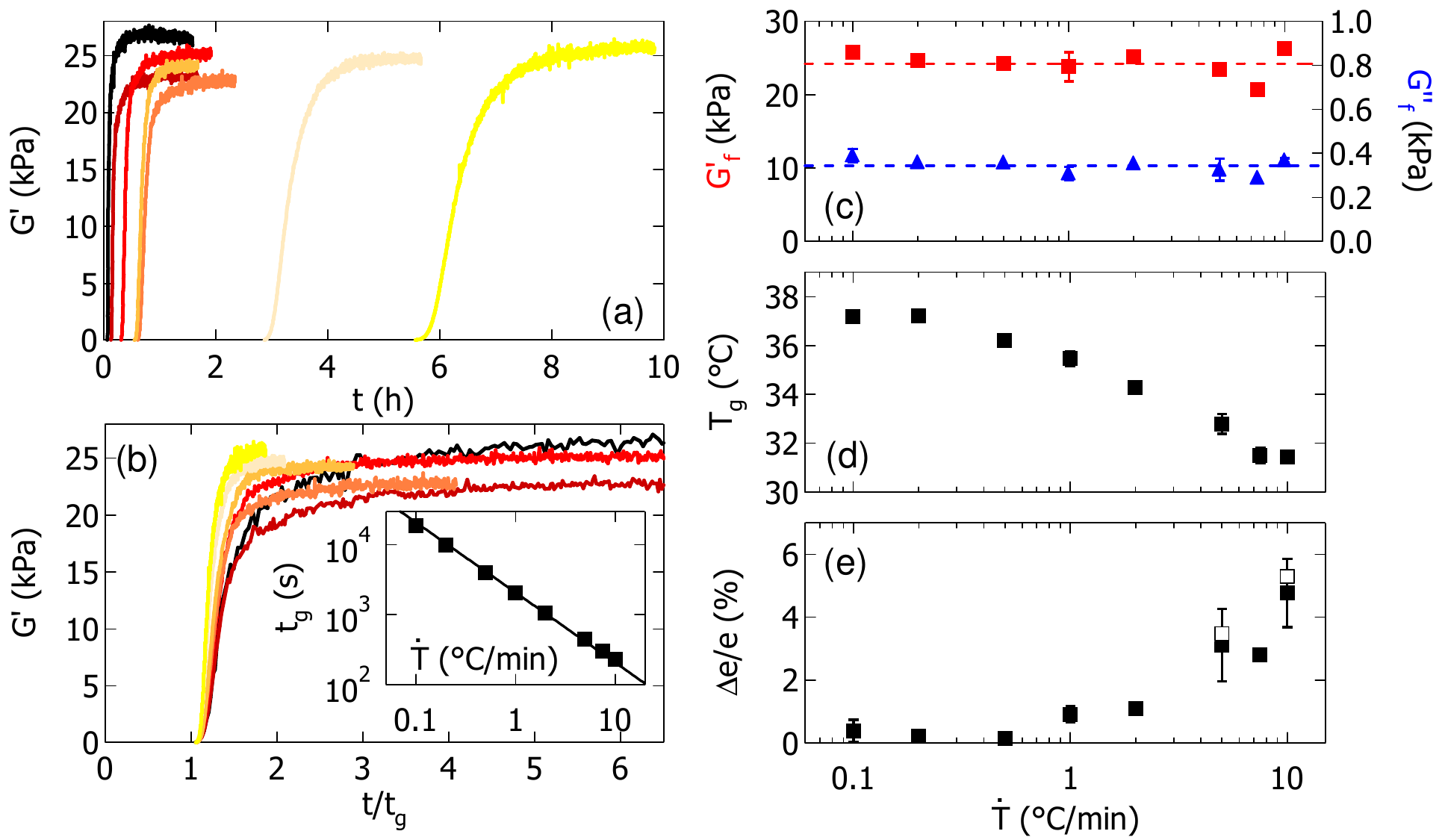}
\caption{\label{fig.5} (Color online) (a) Evolution of the elastic modulus $G'$ vs. time $t$ when decreasing the temperature of a 1.5\% wt. agar solution from $70^{\circ}$C down to $20^{\circ}$C. Colors from black to yellow encode different cooling rates: $\dot {\rm T}=10$, 5, 2, 1, 0.5, 0.2, 0.1$^{\circ}$C/min. (b) Same data set plotted vs. the normalized time $t/t_g$, where $t_g$ is the gelation time defined by the intersection of $G'$ and $G''$.  Inset: gelation time $t_g$ vs. the cooling rate $\dot {\rm T}$. The black line corresponds to the best linear fit of the data in logarithmic scale, which equation is: $t_g\dot {\rm T}=34.9^{\circ}$C. (c) Terminal values $G'_f$ (\textcolor{red}{$\blacksquare$}) and $G''_f$ (\textcolor{blue}{$\blacktriangle$}) of the elastic and viscous moduli respectively vs. the cooling rate $\dot {\rm T}$. The horizontal dashed lines stand for the mean values of the gel viscoelastic moduli: ${\bar G'_f}=(24.2\pm 0.6)$~kPa and ${\bar G''_f}=(340\pm 10$)~Pa. For the sake of clarity, the typical error bar which corresponds to the dispersion of the results obtained by repeating the experiment a few times is indicated only on one data point for $G'_f$. (d) Gelation temperature $T_g$ vs. the cooling rate $\dot {\rm T}$. (e) Relative variation of the gap thickness $e$ vs. the cooling rate $\dot {\rm T}$. Each experiment is performed with a thin layer of oil surrounding the sample. Open squares ($\square$) denote experiments performed with water in the solvent trap and no surrounding oil layer.}%
\end{figure*}

\begin{figure*}[!t]
\includegraphics[width=\linewidth]{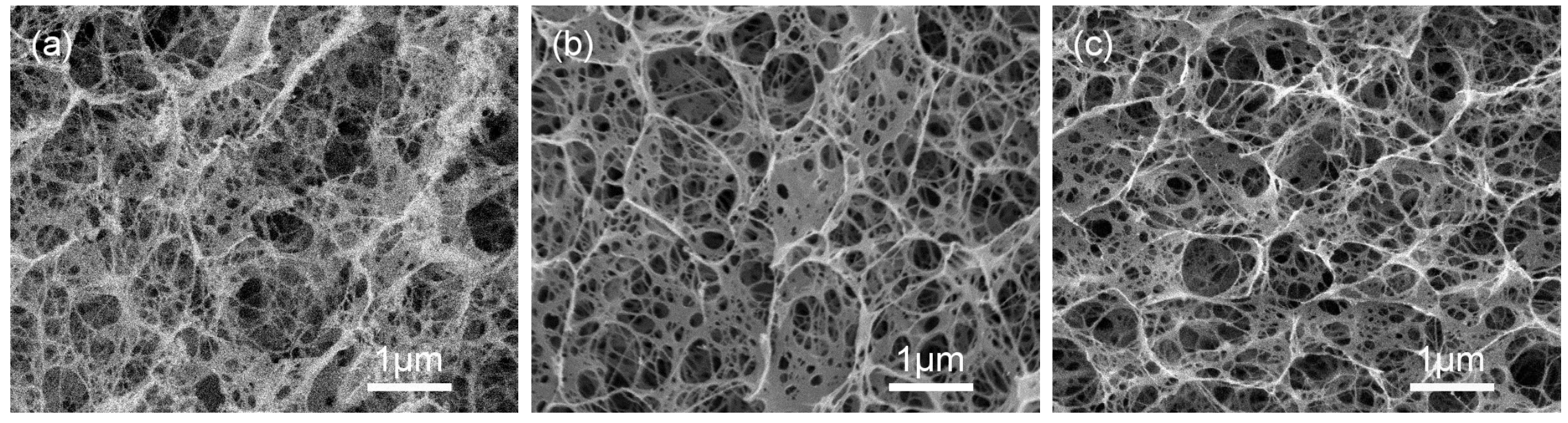}
\caption{\label{fig.12} Cryo-SEM images of 1.5\% wt. agar gels prepared at different cooling rates: (a) $\dot {\rm T}=0.2^{\circ}$C/min, (b) $\dot {\rm T}=1^{\circ}$C/min, and (c) $\dot {\rm T}=10^{\circ}$C/min. In each case, a gel sample is mounted on a pin stub and immersed in liquid nitrogen for about 5~min, before being placed in the low vacuum preparation chamber of a scanning electron microscope (JEOL 6700F). The chamber is cooled down to $T=-90^{\circ}$C and the sample cut in situ with a scalpel. The temperature is increased at a rate of $5^{\circ}$C/min up to $T=-50^{\circ}$C and maintained constant for about 5~min to sublime the water frozen inside the gel. The temperature is decreased back to $T=-85^{\circ}$C. The sample is coated with a nanolayer of gold-palladium, before being cooled down to $T=-160^{\circ}$C and finaly introduced in the observation chamber of the microscope. Samples are imaged in SEI mode, at 5~kV.      
} 
\end{figure*}

Results are displayed in Fig.~\ref{fig.5}. The formation dynamics of the gel strongly depends on $\dot {\rm T} $: gelation occurs sooner for larger cooling rates. More quantitatively, the gelation time $t_g$, defined here as the time at which $G'(t_g) = G''(t_g)$ is inversely proportional to the cooling rate $\dot {\rm T} $ and allows us to rescale the temporal evolution of the elastic modulus into a single mastercurve [Fig.~\ref{fig.5}(b)]. Such a rescaling hints at a unique gelation scenario that is independent of the cooling rate. The gelation temperature, defined as $T_g=T(t_g)$ decreases with the cooling rate: the gelation occurs at $T=37^{\circ}$C at $\dot {\rm T} =0.1^{\circ}$C/min, while the agar sol remains liquid down to 31$^{\circ}$C for a cooling rate of $10^{\circ}$C/min Fig.~\ref{fig.5}(d)]. Interestingly, this evolution does not affect the steady-state properties of the gel at $T=20^{\circ}$C since the terminal values $G'_f$ and $G''_f$ are independent of $\dot {\rm T}$, within error bars [Fig.~\ref{fig.5}(c)]. Scanning cryoelectron microscopy (cryo-SEM) shows that indeed the microstructure of gels prepared at different cooling rates is qualitatively the same (Fig.~\ref{fig.12}) which supports the idea that lower cooling rates only delay the gelation, but do not affect the terminal elastic properties of a 1.5\% wt. agar gel.

This observation strikingly contrasts with results from the literature where agarose gels prepared at constant gap width appear stronger for decreasing cooling rates \cite{Medin:1995,Mohammed:1998}. We attribute such a discrepancy to the partial loss of contact between the gel and the geometry in previous studies, as already discussed in Fig.~\ref{fig.1}(a) and (d). Indeed, the formation of agar gels causes the sample contraction which is robustly observed to become more important for increasing cooling rates, independently of the presence of an oil rim [Fig.~\ref{fig.5}(d)]. The sample contraction increases from about 1\% for $\dot {\rm T} \leq  1^{\circ}$C/min, up to 5\%, at $\dot {\rm T} = 10^{\circ}$C/min. In the latter case, the rapid cooling freezes the microstructure far from its minimum energy state and the relaxation of internal elastic stresses then favors the contraction of the network. For a fixed gap, the contraction is sufficient to trigger a partial loss of contact between the gel and the geometry leading to a lower estimate of the elastic modulus, exactly as evidenced in the experiment reported in Fig.~\ref{fig.1}(a) and (d). Besides, the present observation showing that larger cooling rates lead to larger contractions is compatible with the apparent decrease of the elastic modulus reported for increasing cooling rates in experiments previously reported in the literature at constant gap width. Hence the relevance of the zero normal force protocol, especially for large cooling rates where the gel contraction becomes non-negligible. Finally, one should keep in mind that the value of $\dot {\rm T}$ selects the gelation temperature $T_g$ (roughly estimated from the intersection of $G'$ and $G''$), and does not affect the gel linear rheology determined at 20$^{\circ}$C, at least over two decades of cooling rates from 0.1 to $10^{\circ}$C/min. These two results are not incompatible. Indeed, the elastic modulus of the agar gel scales as a power-law of the concentration in agarose, with an exponent close to 2.2 \cite{Normand:2000}. Therefore a 5\% increase in the agar concentration due to the gap decrease during a rapid cooling results in a 12\% increase in $G'$. The uncertainty in the measurement of $G'$ is of the same order of magnitude as emphasized in section~\ref{Matmet}C, which explains why $G'$ is reported as roughly constant without significant dependence upon the cooling rate.   

\subsection{Impact of the temperature drop}
\label{temperaturedrop}

\begin{figure*}[!t]
\includegraphics[width=0.8\linewidth]{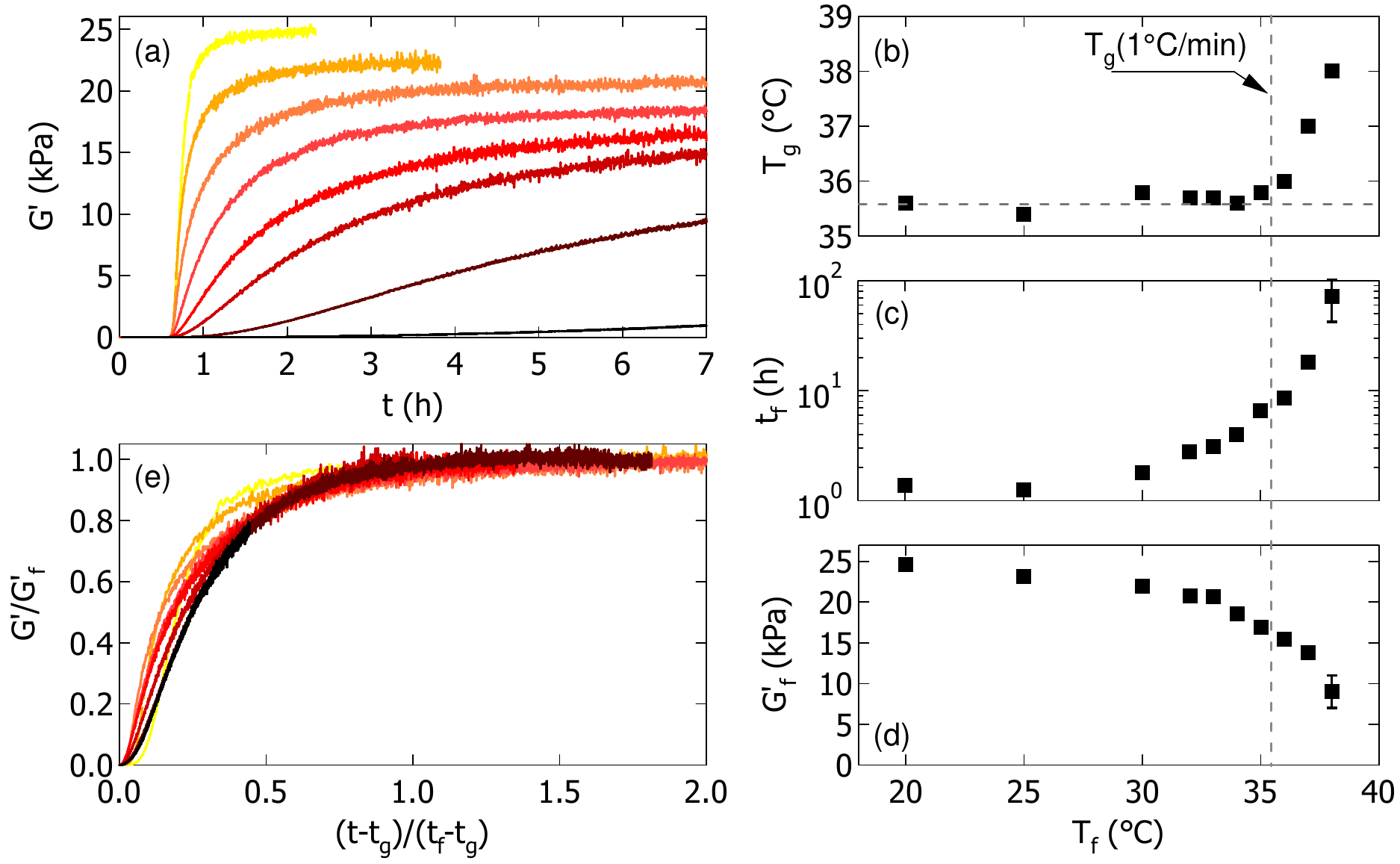}
\caption{\label{fig.6} (Color online) (a)~Evolution of the elastic modulus $G'$ vs. time $t$ while cooling the agar solution at a constant rate  of $\dot {\rm T}=1^{\circ}$C/min, from $T=70^{\circ}$C down to $T_f$. The colors, from yellow to black, stand for different end-of-ramp temperatures: $T_f=20$, 30, 32, 34, 35, 36, 37 and 38$^{\circ}$C. (b)~Gelation temperature $T_g$ as determined by the intersection of $G'$ and $G''$ vs. the terminal temperature $T_f$. (c) Time $t_f$ for the elastic modulus to reach the steady state value vs. the terminal temperature $T_f$ in semilogarithmic scale. (d)~Steady-state value of the elastic modulus $G'_f$ vs. $T_f$. (e) Normalized elastic modulus $G'/G'_f$, where $G'_f$ stands for the steady state value of $G'$, vs. $(t-t_g)/(t_f-t_g)$. Each experiment is performed with a thin layer of oil surrounding the sample.} 
\end{figure*}

The cooling rate is now fixed to $\dot {\rm T}=1^{\circ}$C/min and experiments are performed by varying the final temperature $T_f$ at the end of the ramp so as to quantify the impact of the temperature drop on the gel mechanical properties. The sample is loaded as a liquid into the preheated gap of the rheometer, and the temperature is swept from $70^{\circ}$C down to $T_f$, where $T_f$ is chosen between 20 and $38^{\circ}$C. As illustrated in Fig.~\ref{fig.6}(a), the terminal temperature $T_f$ impacts both the gelation dynamics and the steady-state value of the elastic modulus. 

Imposing a cooling rate $\dot {\rm T}=1^{\circ}$C/min, fixes the gelation temperature to $T_g \simeq 35.5^{\circ}$C [Fig.~\ref{fig.5}(d)]. We will therefore discuss the gelation dynamics whether $T_f$ is larger or smaller than $35.5^{\circ}$C. For $T_f\leq 35.5^{\circ}$C, the gelation always starts at $T_g=35.5^{\circ}$C confirming that the intersection of $G'$ and $G''$ mainly depends on the cooling rate [Fig.~\ref{fig.6}(b)]. 
Moreover, the time $t_f$ for the elastic modulus to reach a steady-state value (within 5\%) increases with $T_f$, which accounts for the increasing duration of the gelation as the terminal temperature is increased [Fig.~\ref{fig.6}(c)]. Surprisingly, for $T_f\geq  35.5^{\circ}$C, the gelation dynamics becomes slower but the gelation is still observed, at least up to $T=38^{\circ}$C. Within this temperature range, the slowdown of the dynamics now results from the increase of both the gelation time $t_g$ and the time $t_f$ to reach the steady state, i.e. for increasing values of $T_f$, the intersection of $G'$ and $G''$ occurs later and once $G'>G''$ it takes more time for $G'$ to reach $G'_f$ [Fig.~\ref{fig.6}(b) and (c)]. Finally, for increasing terminal temperatures $T_f$ one observes the formation of gels with decreasing elastic modulus $G'_f$ [Fig.~\ref{fig.6}(d)]. Such a behavior is observed whether $T_f$ is smaller or larger than $T_g(1^{\circ}\textrm{C/min})\simeq 35.5^{\circ}$C, which suggests that the gel formation scenario is likely to be the same on both sides of $T_g(1^{\circ}\textrm{C/min})$ although the dynamics is strongly slowed down for $T_f>T_g(1^{\circ}\textrm{C/min})$. 

Indeed, here again the various gelation curves $G'(t)$ obtained for different terminal temperature $T_f$ can be rescaled into a single master curve by plotting $G'/G'_f$ versus $(t-t_g)/(t_f-t_g)$ [Fig.~\ref{fig.6}(e)]. The latter function provides the best rescaling among other temporal functions such as $t/t_f$, $t/t_g$, etc. which demonstrates that the duration $t_f-t_g$ separating the beginning of the gelation from the steady state is the relevant unit of time associated with the gelation of 1.5\% wt. agar gels. Furthermore, the existence of a master curve confirms the existence of a single gelation scenario, that is supposedly the same as the one evidenced by the experiments performed at different cooling rates (section~\ref{coolingrate}). This result also indicates that the gel elastic modulus is mostly a function of the terminal temperature $T_f$ reached at the end of the temperature ramp, and that $G'_f(T_f)$ reflects the temperature dependence of a gel with a single microstructure, whether it has been formed above or below 35.5$^{\circ}$C. This point is discussed in more detail in the following paragraph.

\subsection{Impact of a temperature plateau during cooling}

\begin{figure}[h!]
\includegraphics[width=\linewidth]{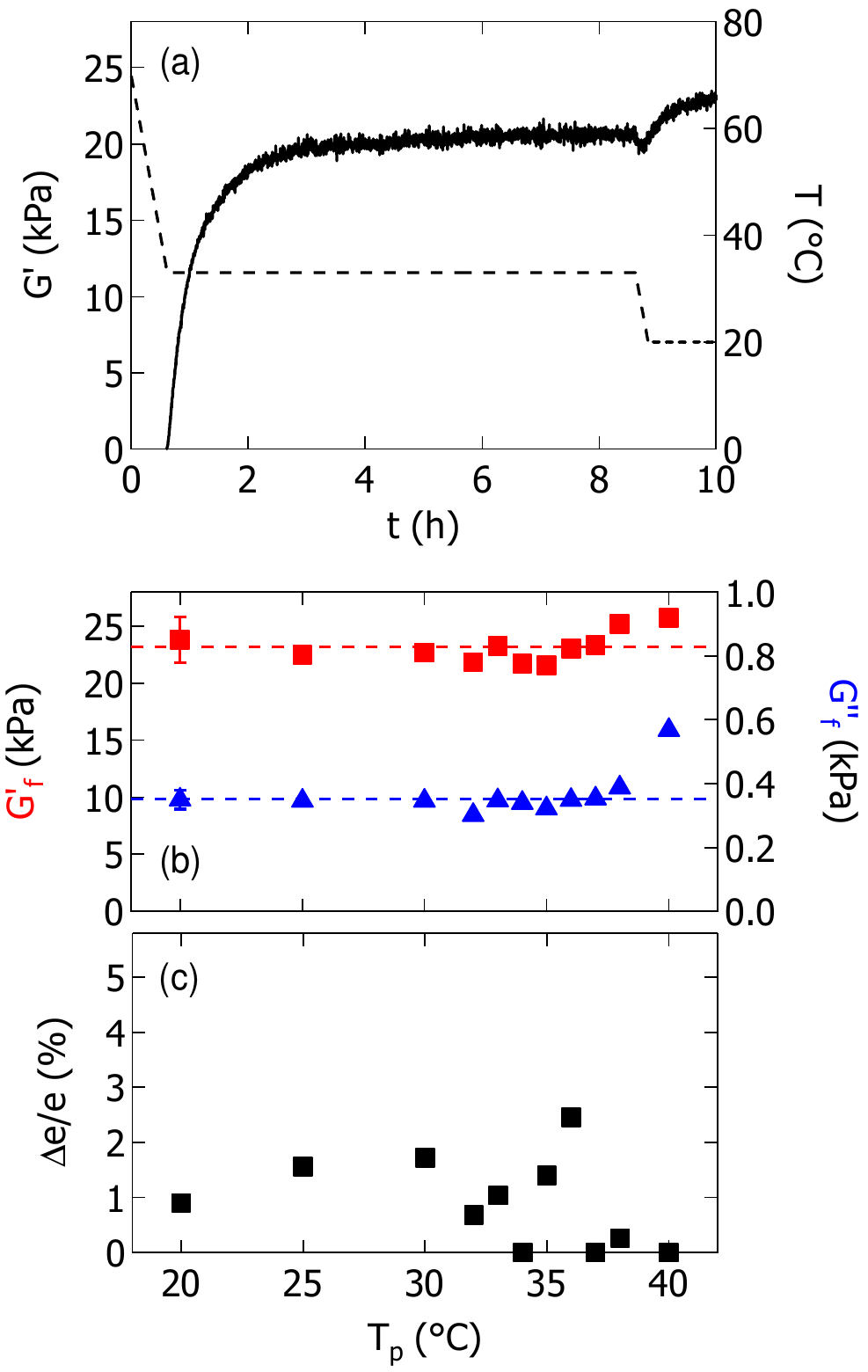}
\caption{\label{fig.7} (Color online) (a) Evolution of the elastic modulus $G'$ (solid line, left axis) and of the temperature $T$ (dashed line, right axis) versus time $t$. The temperature is decreased at a cooling rate $\dot {\rm T}=1^{\circ}$C/min, from $T=70^{\circ}$C down to $T=T_p=33^{\circ}$C. The latter temperature is maintained for 8~h before resuming the cooling down to $T_f=20^{\circ}$C at $\dot {\rm T}=1^{\circ}$C/min. (b) Terminal values $G'_f$ (\textcolor{red}{$\blacksquare$}) and $G''_f$ (\textcolor{blue}{$\blacktriangle$}) of the elastic and viscous moduli vs the temperature $T_p$ of the plateau. For the sake of clarity, the typical error bar determined by repeating the experiment is indicated only on one data point. The horizontal dashed lines stand for the mean values of the gel viscoelastic moduli: ${\bar G'_f}=(23.2\pm 0.5)$~kPa and ${\bar G''_f}=(350\pm 10$)~Pa. (c) Relative variation of the gap vs. $T_p$. Each experiment is performed with a thin layer of oil surrounding the sample.} 
\end{figure}

To determine whether the gels that are obtained at different terminal temperatures on both sides of $T_g(1^{\circ}\textrm{C/min}) \simeq 35.5^{\circ}$C display fundamentally different properties, we perform a last series of experiments. Gelation experiments of 1.5\% wt. agar solutions similar to the ones reported in section~\ref{temperaturedrop} are performed by decreasing the temperature at 1$^{\circ}$C/min from $70^{\circ}$C down to $T=T_p$, where $T_p$ is chosen between 20 and $38^{\circ}$C and maintained for 8~h before resuming the cooling down to $T_f=20^{\circ}$C. The result of a temperature stop at $T_p=33^{\circ}$C is given in Fig.~\ref{fig.7}(a). The elastic modulus reaches a constant value while $T=T_p$ in agreement with the data reported in Fig.~\ref{fig.6}. The subsequent drop of temperature down to $T_f$ leads to a novel increase of $G'$ which terminal value, $G'_f$, is compatible with the value measured by direct cooling from 70$^{\circ}$C to 20$^{\circ}$C, i.e. without any plateau of temperature. The experiments repeated at different values of $T_p$ further support this conclusion: the terminal values reached by $G'$ and $G''$ are independent of $T_p$ [Fig.~\ref{fig.7}(b)], whether the plateau of temperature is performed above or below the gelation temperature $T_f=35.5^{\circ}$C selected by the cooling rate. Such values of $G'_f$ and $G''_f$ are also compatible within error bars with the ones reported in Fig.~\ref{fig.5}(c) for different cooling rates. This result shows that the gelation of a 1.5\% wt. agar solution is mainly controlled by the spinodal demixing expected to take place over shorter timescales than the timescale associated with the cooling process (i.e. 50~min). As a result, the elastic and viscous moduli of the gel formed upon cooling are fixed by the terminal temperature $T_f$ and the thermal history poorly impacts the gel linear properties. Finally, the gel contraction slightly varies from one experiment to the next, but does not show any significant trend with the temperature of the plateau, ranging between 0 and 2\%, with a mean value $\Delta e/e \simeq 1$~\% [Fig.~\ref{fig.7}(c)]. Again, such values justify the use of the zero normal force protocol instead of a constant gap width to determine the gel elastic properties.

\section{Discussion and conclusions}
\label{Discussion}

 In the first part of this work we have shown quantitatively that the zero normal force protocol is more adapted to monitor the gelation dynamics of agar solutions than the traditional constant gap width procedure. Indeed, the gelation of a 1.5\% wt. agar solution results in the sample contraction as evidenced by the large negative normal forces measured in experiments performed at constant gap width. Such a contraction in a cell of constant volume leads to the strain hardening of the gel, which explains the artificial drift of the gel elastic modulus towards ever larger values that is commonly encountered in the literature, even after the gelation is over. Although supplemental observations of the gel formation in a geometry of constant gap width are needed to fully elucidate the local scenario associated with strain hardening and the gel debonding from the walls [see for instance ref. \cite{Arevalo:2015}], we have shown that the zero normal force protocol allows one to determine more reliably the steady-state values of the gel elastic properties. 
Furthermore, we have demonstrated that the zero normal force protocol makes it possible to monitor the gelation with smooth boundary conditions. Therefore, such protocol ensures a good thermal contact between the sample and the bare metallic geometry compared to constant gap width experiments which require the use of sandpaper (or an equivalent coating) to prevent wall slip and/or the gel debonding from the walls. 
As such, the zero normal force protocol appears very promising to investigate more quantitatively the gelation or the solidification dynamics of samples displaying large volume changes during the sol-gel or the liquid-solid transition. In particular, we anticipate that the zero normal force protocol will find relevant applications for cocoa butter rheology \cite{Sonwai:2006,Taylor:2009,Habouzit:2012} and more generally in the rapidly growing field of hydrate slurries \cite{Peixinho:2010,Ahuja:2015} where rheological measurements appear extremely delicate due potentially to the relative dilation of the sample compared to that of the geometry under complex thermal protocols. 

Moreover, we have unraveled two subtle artifacts related to the contact of an aqueous solution with metallic boundary conditions that may lead to a premature increase of the elastic modulus long before the start of the gelation. On the one hand, the corrosion of duralumin plates by the oxidative agar sol may enrich the air/solution or the oil/solution interface with nanoparticles. Such nanoparticles increase the elasticity of the contact line before the start of the gelation, leading to an artificial increase of the elastic modulus, with $G'>G''$. Note that the latter issue is specific to duralumin and is not observed for plates made of stainless steel. On the other hand, the oil layer traditionally added around the sample to prevent evaporation may also affect the rheological measurements while the sample is still liquid. The oil may slowly invade the gap leading to a premature increase of $G'$. The role of the plates wetting properties on the oil invasion scenario remains to be investigated. 

Finally, in the second part of this article, we have applied the zero normal force protocol to determine the influence of thermal history on the linear rheology of a 1.5\% wt. agar gel containing 1\% wt. of agarose. Our study demonstrates that the gel elastic properties are mostly controlled by the terminal temperature at the end of the cooling process, and that contrarily to gels with higher agarose content \cite{Aymard:2001}, neither the cooling rate nor the thermal path affect the gel mechanical properties. Such a result suggests that the local gelation scenario is mostly controlled by the rapid spinodal demixing of the agar solution in contrast with the complex formation scenario reported in \cite{Manno:1999} on a 2\% wt agarose gel. As a practical consequence, the thermal parameters allow one to modulate the gel formation dynamics. Namely, the crossover between the elastic and viscous modulus is selected by the cooling rate ${\rm \dot T}$, while for a given value of ${\rm \dot T}$, the choice of the terminal temperature relative to the temperature at which the crossover of $G'$ and $G''$ takes place makes it possible to strongly slow down the gelation. These results should help in the optimization of manufacturing lines of agar-based products such as Petri dishes or specific electrophoretic gels, for which the cooling rate and the delay to complete gelation are the most constraining factors regarding the production rate. More generally, the zero normal force procedure should help performing more reliable rheological investigation of non-isochoric processes.

\begin{acknowledgments}
This work received funding from BioM\'erieux and the ANRT under the CIFRE program, Grant Agreement No.~112972. The authors acknowledge H.~Saadaoui for AFM measurements of the surface roughness, J.-M. Olive for polishing the duralumin plates, P.~Legros for his help with the cryo-SEM experiments, as well as Y.~Amarouch\`ene, J.-B. Salmon, F.~ Villeval and E.~Laurichesse for stimulating discussions, and two anonymous referees for constructive comments on our manuscript. T.D. acknowledges funding from the CNRS through the``Theoretical Physics and Interfaces" PEPS scheme, project ``ComplexWall". 
\end{acknowledgments}

\begin{figure}[b!]
\includegraphics[width=\linewidth]{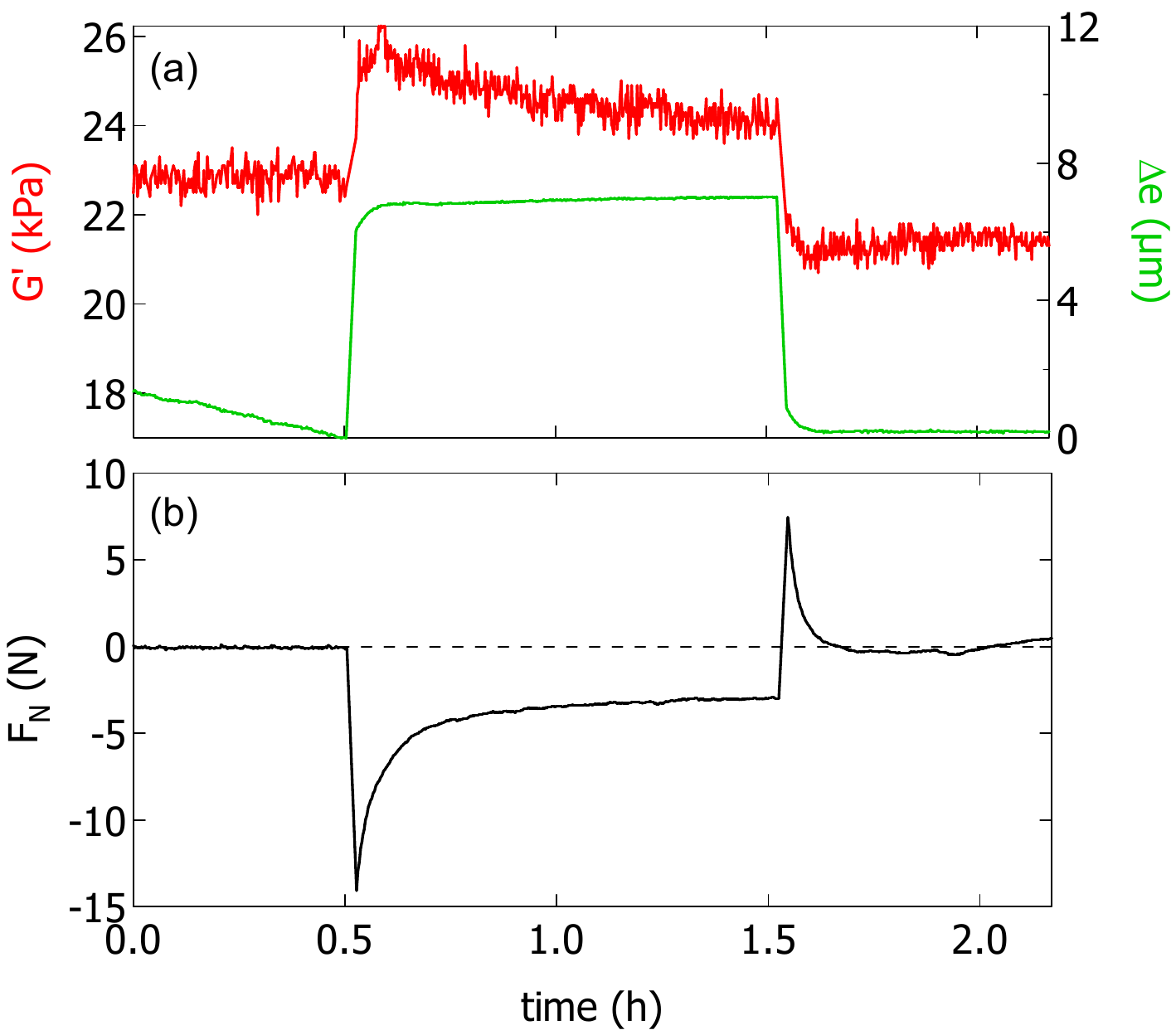}
\caption{\label{fig.sup1bis} (Color online) (a) Temporal response of the gel elastic modulus $G'$ to an increase of the gap width of $\Delta e=6~\mu$m triggered at $t=0.5$~h and suppressed after 1~h. $\Delta e$ denotes the excess gap width above the minimum value reached at t=0.5h, at the end of the gelation process under controlled normal force, and taken as an arbitrary reference. (b) Temporal evolution of the normal force $F_N$ concomitant to the gel response.} 
\end{figure}

\begin{figure*}[t!]
\includegraphics[width=0.8\linewidth]{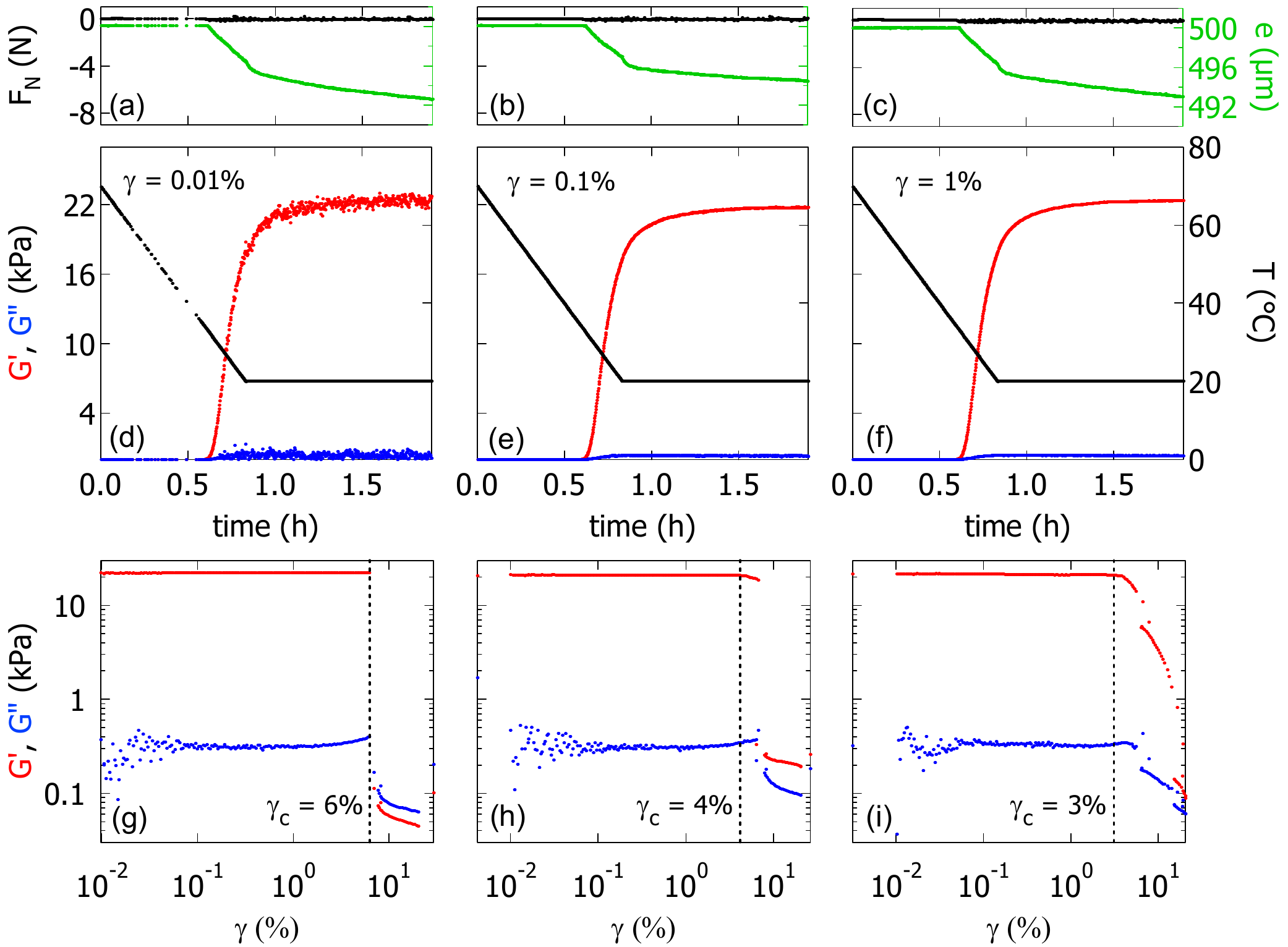}
\caption{\label{fig.sup2} (Color online) Temporal evolution of the normal force $F_N$ and the gap width $e$ (a)--(c) together with the elastic and viscous moduli (d)--(e) during the gelation of a 1.5\%~wt. agar solution induced by decreasing the temperature from $T=70^{\circ}$C to 20$^{\circ}$C at a cooling rate $\dot {\rm T}=1^{\circ}$C/min. The gelation experiment is repeated three times using the zero normal force protocol and applying different constant strain amplitudes: $\gamma=0.01$\%, 0.1\% and 1\% in respectively the first, second and third column. Terminal values of the gel elastic modulus are respectively: 22.3, 21.8 and 22.3~kPa, and compatible within error bars. For all the experiments the oscillation frequency is 1Hz. (g)-(i) Strain sweeps experiments performed at 1Hz on the three gels obtained in (d), (e) and (f) after two hours, once they are fully gelified. The vertical dashed line denotes the critical strain $\gamma_c$ above which $G'$ decreases.} 
\end{figure*}

\section*{Appendix}

The first part of the appendix provides a proof that the slow increase of the elastic modulus reported in Fig.~\ref{fig.1}(e) during a gelation experiment performed at constant gap width, is indeed due to the strain hardening of the sample. We have performed the following supplemental test: an agar gel is prepared between parallel plates (initial gap $e_0=$500~$\mu$m) using the zero normal force protocol, by decreasing the temperature from $70^{\circ}$C down to $20^{\circ}$C at $1^{\circ}$C/min. At the end of the gelation process, the gap has decreased by 1\% and the normal force is still equal to zero, in agreement with the results reported in Fig.~\ref{fig.1}(c) and (f) in the main text. We then switch the rheometer to a controlled gap mode and increase the gap width by $\Delta e=6~\mu$m at $0.1~\mu$m/s, which corresponds to a vertical extension of the gel of about 1\%. In response, the gel elastic modulus reported in Fig.~\ref{fig.sup1bis}(a) shows a brutal increase of about 10\% from 23~kPa to 25~kPa, while the normal force concomitantly becomes negative and relaxes towards $F_N \simeq -3$N [Fig.~\ref{fig.sup1bis}(a)] which indicates that the gel is pulling on the upper plate. These results show that an increase of the gap width triggers an increase of the gel elastic modulus: the gel experiences strain hardening under extension. We thus conclude that the contraction of the sample at constant gap width also leads to the strain hardening of the sample, which explains the ever increasing values of $G'$ reported during gelations performed at constant gap width [Fig.~\ref{fig.1}(e)]. Moreover, when the gap is brought back to its initial position after 1~h, the normal force relaxes back to zero, while $G'$ decreases to a value slightly lower than the initial one (Fig.~\ref{fig.sup1bis}). Such a discrepancy is likely due to a partial loss of contact between the gel and the plates, but remains to be investigated in more details.
\begin{figure*}[t!]
\includegraphics[width=0.7\linewidth]{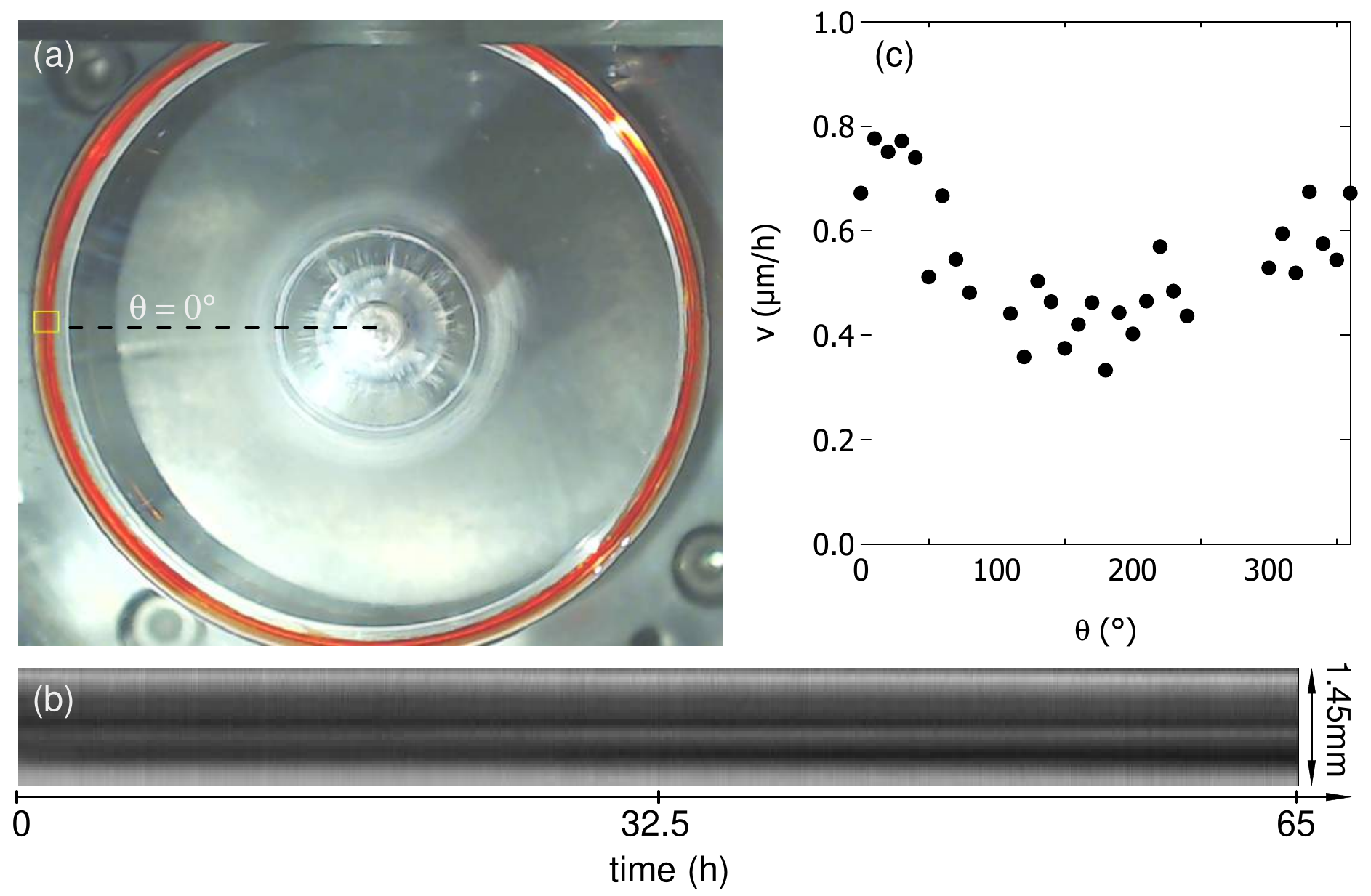}
\caption{\label{fig.sup1} (Color online) (a) Picture of the parallel plate geometry shot from below. Both the upper and bottom plates are made of transparent PMMA. The gap is filled with distilled water and surrounded by an oil rim of about 1~mm thick and stained in red with a fat soluble dye. The scale is set by the plate diameter of 40~mm. (b) Spatiotemporal diagram of the oil/water interface as a function of time and computed in the region of interest (ROI) emphasized by the yellow rectangle in (a). (c) Radial velocity $v$ of the water/oil interface computed over 65~h vs. the angular position $\theta$ of the ROI. The ROI reference angular position ($\theta=0^{\circ}$) is indicated in (a) and the angle $\theta$ is counted positively in the counterclockwise direction.} 
\end{figure*}

The second part of the appendix is dedicated to the impact of the strain amplitude during the gelation experiments. We have repeated the gelation experiment reported in the main text [Fig.~\ref{fig.1}(f)] at different strain amplitudes of constant value: $\gamma=0.01$, 0.1 and 1\% while applying the zero normal force protocol. The results displayed in Fig.~\ref{fig.sup2} show quantitatively the same terminal values of the elastic modulus [$G'_f\simeq 22$~kPa in Fig.~\ref{fig.sup2}(d)--(f)] and the same sample contraction [$\Delta e/e \simeq 1$\% in Fig.~\ref{fig.sup2}(a)--(c)]. Therefore, as long as the strain amplitude remains lower than 1\%, the stress applied during the gelation performed with the zero normal force protocol does not impact the gel formation dynamics nor the steady-state parameters of the gel. Yet the strain applied during the gelation does matter. Indeed, we have performed a strain sweep on the three gels which have been exposed to different oscillatory amplitudes during their gelation [Fig.~\ref{fig.sup2}(g)--(i)]. For all the experiments, we observe that G' decreases above a critical strain $\gamma_c$ which is interpreted as the progressive debonding of the gel from the plate, as the gel remains intact and no macroscopic crack are visible at the end of the strain sweep. First, the critical strain $\gamma_c$ above which $G'$ decreases as a result of the gel debonding is always larger than 1\%. This result further justifies the strain adapted protocol introduced in section~\ref{BC} of the main text, which consists in progressively decreasing the strain amplitude applied to measure the viscoelastic modulus during the gelation. Second and as a remarkable result, the critical strain $\gamma_c$ above which $G'$ decreases depends on the strain amplitude applied during the gelation. Gels which formation has been monitored with a smaller amplitude, maintain contact with the plates up to a larger value of $\gamma_c$. This observation indicates that small strain amplitude as low as 0.01\% during the gelation are required for an optimal adhesion between the sample and the plates. Finally, the evolution of the viscoelastic modulus in the vicinity of the critical deformation $\gamma_c$ strongly differs depending on the strain amplitude applied during the gel formation. The gel which has been prepared at $\gamma=0.01$\% shows a brutal drop in $G'$ during the strain sweep [Fig.~\ref{fig.sup2}(g)] which hints at an abrupt detachment from the plates. Gels which formation have been monitored under larger strains ($\gamma=0.1$ and 1\%) display a much smoother decrease in $G'$ above $\gamma_c$ [Fig.~\ref{fig.sup2}(h) and (i)] which could correspond to a more progressive contact loss from the plate. This last point certainly deserves more investigation. 

The third and last part of the appendix concerns the displacement of the peripheral oil/water interface sandwiched between two transparent PMMA plates, as illustrated in Fig.~\ref{fig.4b} in the main text. Let us recall that the gap is filled with distilled water and that the surrounding oil rim is stained in red with a fat soluble dye. We use a spatiotemporal analysis to measure the velocity of the interface. A picture of the entire geometry shot from below is reported in Fig.~\ref{fig.sup1}(a). A rectangular region of interest ($1.45$~mm$\times 1.2$~mm, i.e. 39$\times$32 pixels$^2$) is pictured as a yellow rectangle in Fig.~\ref{fig.sup1}(a) and defines the reference angular position $\theta=0^{\circ}$ with respect to the plate rotation axis. The temporal evolution of the gray level along the oil/water interface is plotted as a spatiotemporal diagram in Fig.~\ref{fig.sup1}(b) which allows us to measure the inclination of the grey level texture through a sub-pixel digital image correlation and detect a displacement of about one pixel over a duration of 65~h! The latter analysis applied every 10$^{\circ}$ along the periphery of the oil/water interface reveals that the oil moves systematically towards the rotation axis at a velocity $v$ which depends on the angular position $\theta$ and ranges between 0.4 and 0.8~$\mu$m/h [Fig.~\ref{fig.sup1}(c)].

\section*{References}


\begin{thebibliography}{79}%
\makeatletter
\providecommand \@ifxundefined [1]{%
 \@ifx{#1\undefined}
}%
\providecommand \@ifnum [1]{%
 \ifnum #1\expandafter \@firstoftwo
 \else \expandafter \@secondoftwo
 \fi
}%
\providecommand \@ifx [1]{%
 \ifx #1\expandafter \@firstoftwo
 \else \expandafter \@secondoftwo
 \fi
}%
\providecommand \natexlab [1]{#1}%
\providecommand \enquote  [1]{``#1''}%
\providecommand \bibnamefont  [1]{#1}%
\providecommand \bibfnamefont [1]{#1}%
\providecommand \citenamefont [1]{#1}%
\providecommand \href@noop [0]{\@secondoftwo}%
\providecommand \href [0]{\begingroup \@sanitize@url \@href}%
\providecommand \@href[1]{\@@startlink{#1}\@@href}%
\providecommand \@@href[1]{\endgroup#1\@@endlink}%
\providecommand \@sanitize@url [0]{\catcode `\\12\catcode `\$12\catcode
  `\&12\catcode `\#12\catcode `\^12\catcode `\_12\catcode `\%12\relax}%
\providecommand \@@startlink[1]{}%
\providecommand \@@endlink[0]{}%
\providecommand \url  [0]{\begingroup\@sanitize@url \@url }%
\providecommand \@url [1]{\endgroup\@href {#1}{\urlprefix }}%
\providecommand \urlprefix  [0]{URL }%
\providecommand \Eprint [0]{\href }%
\providecommand \doibase [0]{http://dx.doi.org/}%
\providecommand \selectlanguage [0]{\@gobble}%
\providecommand \bibinfo  [0]{\@secondoftwo}%
\providecommand \bibfield  [0]{\@secondoftwo}%
\providecommand \translation [1]{[#1]}%
\providecommand \BibitemOpen [0]{}%
\providecommand \bibitemStop [0]{}%
\providecommand \bibitemNoStop [0]{.\EOS\space}%
\providecommand \EOS [0]{\spacefactor3000\relax}%
\providecommand \BibitemShut  [1]{\csname bibitem#1\endcsname}%
\let\auto@bib@innerbib\@empty
\bibitem [{\citenamefont {Mezger}(2014)}]{Mezger:2014}%
  \BibitemOpen
  \bibfield  {author} {\bibinfo {author} {\bibfnamefont {T.~G.}\ \bibnamefont
  {Mezger}},\ }\href@noop {} {\emph {\bibinfo {title} {The Rheology
  Handbook}}},\ edited by\ \bibinfo {editor} {\bibfnamefont {E.}~\bibnamefont
  {Coatings}}\ (\bibinfo  {publisher} {Vincentz Network; 4 edition},\ \bibinfo
  {year} {2014})\BibitemShut {NoStop}%
\bibitem [{\citenamefont {Senff}\ and\ \citenamefont
  {Richtering}(1999)}]{Senff:1999}%
  \BibitemOpen
  \bibfield  {author} {\bibinfo {author} {\bibfnamefont {H.}~\bibnamefont
  {Senff}}\ and\ \bibinfo {author} {\bibfnamefont {W.}~\bibnamefont
  {Richtering}},\ }\bibfield  {title} {\enquote {\bibinfo {title} {Temperature
  sensitive microgel suspensions: Colloidal phase behavior and rheology of soft
  spheres},}\ }\href@noop {} {\bibfield  {journal} {\bibinfo  {journal}
  {Journal of Chemical Physics}\ }\textbf {\bibinfo {volume} {111}},\ \bibinfo
  {pages} {1705--1711} (\bibinfo {year} {1999})}\BibitemShut {NoStop}%
\bibitem [{\citenamefont {Kapnistos}\ \emph {et~al.}(2000)\citenamefont
  {Kapnistos}, \citenamefont {Vlassopoulos}, \citenamefont {Mortensen},
  \citenamefont {Fleisher},\ and\ \citenamefont {Roovers}}]{Kapnistos:2000}%
  \BibitemOpen
  \bibfield  {author} {\bibinfo {author} {\bibfnamefont {M.}~\bibnamefont
  {Kapnistos}}, \bibinfo {author} {\bibfnamefont {D.}~\bibnamefont
  {Vlassopoulos}}, \bibinfo {author} {\bibfnamefont {G.~F.~F.}\ \bibnamefont
  {Mortensen}}, \bibinfo {author} {\bibfnamefont {G.}~\bibnamefont {Fleisher}},
  \ and\ \bibinfo {author} {\bibfnamefont {J.}~\bibnamefont {Roovers}},\
  }\bibfield  {title} {\enquote {\bibinfo {title} {Reversible thermal glelation
  in soft spheres},}\ }\href@noop {} {\bibfield  {journal} {\bibinfo  {journal}
  {Phys. Rev. Lett.}\ }\textbf {\bibinfo {volume} {85}},\ \bibinfo {pages}
  {4072--4075} (\bibinfo {year} {2000})}\BibitemShut {NoStop}%
\bibitem [{\citenamefont {Zhang}\ \emph {et~al.}(2009)\citenamefont {Zhang},
  \citenamefont {N.Xu}, \citenamefont {Chen}, \citenamefont {Yunker},
  \citenamefont {Alsayed}, \citenamefont {Aptowicz}, \citenamefont {Habdas},
  \citenamefont {Liu}, \citenamefont {Nagel},\ and\ \citenamefont
  {Yodh}}]{Zhang:2009}%
  \BibitemOpen
  \bibfield  {author} {\bibinfo {author} {\bibfnamefont {Z.}~\bibnamefont
  {Zhang}}, \bibinfo {author} {\bibnamefont {N.Xu}}, \bibinfo {author}
  {\bibfnamefont {D.}~\bibnamefont {Chen}}, \bibinfo {author} {\bibfnamefont
  {P.}~\bibnamefont {Yunker}}, \bibinfo {author} {\bibfnamefont
  {A.}~\bibnamefont {Alsayed}}, \bibinfo {author} {\bibfnamefont
  {K.}~\bibnamefont {Aptowicz}}, \bibinfo {author} {\bibfnamefont
  {P.}~\bibnamefont {Habdas}}, \bibinfo {author} {\bibfnamefont
  {A.}~\bibnamefont {Liu}}, \bibinfo {author} {\bibfnamefont {S.}~\bibnamefont
  {Nagel}}, \ and\ \bibinfo {author} {\bibfnamefont {A.}~\bibnamefont {Yodh}},\
  }\bibfield  {title} {\enquote {\bibinfo {title} {Thermal vestige of the
  zero-temperature jamming transition},}\ }\href@noop {} {\bibfield  {journal}
  {\bibinfo  {journal} {Nature}\ }\textbf {\bibinfo {volume} {459}},\ \bibinfo
  {pages} {230--233} (\bibinfo {year} {2009})}\BibitemShut {NoStop}%
\bibitem [{\citenamefont {Wang}\ \emph {et~al.}(2014)\citenamefont {Wang},
  \citenamefont {Wu}, \citenamefont {Zhu}, \citenamefont {Liu},\ and\
  \citenamefont {Zhang}}]{Wang:2014b}%
  \BibitemOpen
  \bibfield  {author} {\bibinfo {author} {\bibfnamefont {H.}~\bibnamefont
  {Wang}}, \bibinfo {author} {\bibfnamefont {X.}~\bibnamefont {Wu}}, \bibinfo
  {author} {\bibfnamefont {Z.}~\bibnamefont {Zhu}}, \bibinfo {author}
  {\bibfnamefont {C.}~\bibnamefont {Liu}}, \ and\ \bibinfo {author}
  {\bibfnamefont {Z.}~\bibnamefont {Zhang}},\ }\bibfield  {title} {\enquote
  {\bibinfo {title} {Revisit to phase diagram of poly(n-isopropylacrylamide)
  microgel suspensions by mechanical spectroscopy},}\ }\href@noop {} {\bibfield
   {journal} {\bibinfo  {journal} {The Journal of Chemical Physics}\ }\textbf
  {\bibinfo {volume} {140}},\ \bibinfo {pages} {024908} (\bibinfo {year}
  {2014})}\BibitemShut {NoStop}%
\bibitem [{\citenamefont {Kan\'e}, \citenamefont {Djabourov},\ and\
  \citenamefont {Volle}(2004)}]{Kane:2004}%
  \BibitemOpen
  \bibfield  {author} {\bibinfo {author} {\bibfnamefont {M.}~\bibnamefont
  {Kan\'e}}, \bibinfo {author} {\bibfnamefont {M.}~\bibnamefont {Djabourov}}, \
  and\ \bibinfo {author} {\bibfnamefont {J.-L.}\ \bibnamefont {Volle}},\
  }\bibfield  {title} {\enquote {\bibinfo {title} {Rheology and structure of
  waxy crude oils in quiescent and under shearing conditions},}\ }\href@noop {}
  {\bibfield  {journal} {\bibinfo  {journal} {Fuel}\ }\textbf {\bibinfo
  {volume} {83}},\ \bibinfo {pages} {1591--1605} (\bibinfo {year}
  {2004})}\BibitemShut {NoStop}%
\bibitem [{\citenamefont {Visintin}\ \emph {et~al.}(2005)\citenamefont
  {Visintin}, \citenamefont {Lapasin}, \citenamefont {Vignati}, \citenamefont
  {D'Antona},\ and\ \citenamefont {Lockhart}}]{Visintin:2005}%
  \BibitemOpen
  \bibfield  {author} {\bibinfo {author} {\bibfnamefont {R.}~\bibnamefont
  {Visintin}}, \bibinfo {author} {\bibfnamefont {R.}~\bibnamefont {Lapasin}},
  \bibinfo {author} {\bibfnamefont {E.}~\bibnamefont {Vignati}}, \bibinfo
  {author} {\bibfnamefont {P.}~\bibnamefont {D'Antona}}, \ and\ \bibinfo
  {author} {\bibfnamefont {T.}~\bibnamefont {Lockhart}},\ }\bibfield  {title}
  {\enquote {\bibinfo {title} {Rheological behavior and structural
  interpretation of waxy crude oil gels},}\ }\href@noop {} {\bibfield
  {journal} {\bibinfo  {journal} {Langmuir}\ }\textbf {\bibinfo {volume}
  {21}},\ \bibinfo {pages} {6240--6249} (\bibinfo {year} {2005})}\BibitemShut
  {NoStop}%
\bibitem [{\citenamefont {te~Nijenhuis}(1997)}]{Nijenhuis:1997}%
  \BibitemOpen
  \bibfield  {author} {\bibinfo {author} {\bibfnamefont {K.}~\bibnamefont
  {te~Nijenhuis}},\ }\href@noop {} {\emph {\bibinfo {title} {Thermoreversible
  networks: viscoelastic properties and structure of gels}}},\ edited by\
  \bibinfo {editor} {\bibnamefont {Springer-Verlag}}\ (\bibinfo  {publisher}
  {Springer Berlin Heidelberg},\ \bibinfo {year} {1997})\BibitemShut {NoStop}%
\bibitem [{\citenamefont {Stephen}, \citenamefont {Phillips},\ and\
  \citenamefont {Williams}(2006)}]{FPTA:2006}%
  \BibitemOpen
  \bibinfo {editor} {\bibfnamefont {A.}~\bibnamefont {Stephen}}, \bibinfo
  {editor} {\bibfnamefont {G.}~\bibnamefont {Phillips}}, \ and\ \bibinfo
  {editor} {\bibfnamefont {P.}~\bibnamefont {Williams}},\ eds.,\ \href@noop {}
  {\emph {\bibinfo {title} {Food Polysaccharides and Their Applications}}}\
  (\bibinfo  {publisher} {CRC Press, Taylor and Francis Group -- 2nd ed.},\
  \bibinfo {year} {2006})\BibitemShut {NoStop}%
\bibitem [{\citenamefont {Souguir}\ \emph {et~al.}(2015)\citenamefont
  {Souguir}, \citenamefont {Ronsin}, \citenamefont {Caroli},\ and\
  \citenamefont {Baumberger}}]{Souguir:2015}%
  \BibitemOpen
  \bibfield  {author} {\bibinfo {author} {\bibfnamefont {H.}~\bibnamefont
  {Souguir}}, \bibinfo {author} {\bibfnamefont {O.}~\bibnamefont {Ronsin}},
  \bibinfo {author} {\bibfnamefont {C.}~\bibnamefont {Caroli}}, \ and\ \bibinfo
  {author} {\bibfnamefont {T.}~\bibnamefont {Baumberger}},\ }\bibfield  {title}
  {\enquote {\bibinfo {title} {Two-step build-up of a thermoreversible polymer
  network: From early local to late collective dynamics},}\ }\href@noop {}
  {\bibfield  {journal} {\bibinfo  {journal} {Phys. Rev. E}\ }\textbf {\bibinfo
  {volume} {91}},\ \bibinfo {pages} {042305} (\bibinfo {year}
  {2015})}\BibitemShut {NoStop}%
\bibitem [{\citenamefont {Helgeson}\ \emph {et~al.}(2012)\citenamefont
  {Helgeson}, \citenamefont {Moran}, \citenamefont {An},\ and\ \citenamefont
  {Doyle}}]{Helgeson:2012}%
  \BibitemOpen
  \bibfield  {author} {\bibinfo {author} {\bibfnamefont {M.}~\bibnamefont
  {Helgeson}}, \bibinfo {author} {\bibfnamefont {S.}~\bibnamefont {Moran}},
  \bibinfo {author} {\bibfnamefont {H.}~\bibnamefont {An}}, \ and\ \bibinfo
  {author} {\bibfnamefont {P.}~\bibnamefont {Doyle}},\ }\bibfield  {title}
  {\enquote {\bibinfo {title} {Mesoporous organohydrogels from thermogelling
  photocrosslinkable nanoemulsions},}\ }\href@noop {} {\bibfield  {journal}
  {\bibinfo  {journal} {Nature Materials}\ }\textbf {\bibinfo {volume} {11}},\
  \bibinfo {pages} {344--352} (\bibinfo {year} {2012})}\BibitemShut {NoStop}%
\bibitem [{\citenamefont {Helgeson}\ \emph {et~al.}(2014)\citenamefont
  {Helgeson}, \citenamefont {Gao}, \citenamefont {Moran}, \citenamefont {Lee},
  \citenamefont {Godfrin}, \citenamefont {Tripathi}, \citenamefont {Bose},\
  and\ \citenamefont {Doyle}}]{Helgeson:2014}%
  \BibitemOpen
  \bibfield  {author} {\bibinfo {author} {\bibfnamefont {M.}~\bibnamefont
  {Helgeson}}, \bibinfo {author} {\bibfnamefont {Y.}~\bibnamefont {Gao}},
  \bibinfo {author} {\bibfnamefont {S.}~\bibnamefont {Moran}}, \bibinfo
  {author} {\bibfnamefont {J.}~\bibnamefont {Lee}}, \bibinfo {author}
  {\bibfnamefont {M.}~\bibnamefont {Godfrin}}, \bibinfo {author} {\bibfnamefont
  {A.}~\bibnamefont {Tripathi}}, \bibinfo {author} {\bibfnamefont
  {A.}~\bibnamefont {Bose}}, \ and\ \bibinfo {author} {\bibfnamefont
  {P.}~\bibnamefont {Doyle}},\ }\bibfield  {title} {\enquote {\bibinfo {title}
  {Homogeneous percolation versus arrested phase separation in
  attractively-driven nanoemulsion colloidal gels},}\ }\href@noop {} {\bibfield
   {journal} {\bibinfo  {journal} {Soft Matter}\ }\textbf {\bibinfo {volume}
  {10}},\ \bibinfo {pages} {3122--3133} (\bibinfo {year} {2014})}\BibitemShut
  {NoStop}%
\bibitem [{\citenamefont {Zhang}\ \emph {et~al.}(2001)\citenamefont {Zhang},
  \citenamefont {Yoshimura}, \citenamefont {Nishinari}, \citenamefont
  {Williams}, \citenamefont {Foster},\ and\ \citenamefont
  {Norton}}]{Zhang:2001}%
  \BibitemOpen
  \bibfield  {author} {\bibinfo {author} {\bibfnamefont {H.}~\bibnamefont
  {Zhang}}, \bibinfo {author} {\bibfnamefont {M.}~\bibnamefont {Yoshimura}},
  \bibinfo {author} {\bibfnamefont {K.}~\bibnamefont {Nishinari}}, \bibinfo
  {author} {\bibfnamefont {M.}~\bibnamefont {Williams}}, \bibinfo {author}
  {\bibfnamefont {T.}~\bibnamefont {Foster}}, \ and\ \bibinfo {author}
  {\bibfnamefont {I.}~\bibnamefont {Norton}},\ }\bibfield  {title} {\enquote
  {\bibinfo {title} {Gelation behaviour o fkonjac glucomannan with different
  molecular weights},}\ }\href@noop {} {\bibfield  {journal} {\bibinfo
  {journal} {Biopolymers}\ }\textbf {\bibinfo {volume} {59}},\ \bibinfo {pages}
  {38--50} (\bibinfo {year} {2001})}\BibitemShut {NoStop}%
\bibitem [{\citenamefont {Gong}\ and\ \citenamefont {Osada}(2010)}]{Gong:2010}%
  \BibitemOpen
  \bibfield  {author} {\bibinfo {author} {\bibfnamefont {J.}~\bibnamefont
  {Gong}}\ and\ \bibinfo {author} {\bibfnamefont {Y.}~\bibnamefont {Osada}},\
  }\bibfield  {title} {\enquote {\bibinfo {title} {Soft and wet materials: From
  hydrogels to biotissues},}\ }\href@noop {} {\bibfield  {journal} {\bibinfo
  {journal} {Advances in Polymer Science}\ }\textbf {\bibinfo {volume} {236}},\
  \bibinfo {pages} {203--246} (\bibinfo {year} {2010})}\BibitemShut {NoStop}%
\bibitem [{\citenamefont {Nguyen}\ and\ \citenamefont
  {Lee}(2010)}]{Nguyen:2010}%
  \BibitemOpen
  \bibfield  {author} {\bibinfo {author} {\bibfnamefont {M.~K.}\ \bibnamefont
  {Nguyen}}\ and\ \bibinfo {author} {\bibfnamefont {D.}~\bibnamefont {Lee}},\
  }\bibfield  {title} {\enquote {\bibinfo {title} {Injectable biodegradable
  hydrogels},}\ }\href@noop {} {\bibfield  {journal} {\bibinfo  {journal}
  {Macromol. Biosci.}\ }\textbf {\bibinfo {volume} {10}},\ \bibinfo {pages}
  {563--579} (\bibinfo {year} {2010})}\BibitemShut {NoStop}%
\bibitem [{\citenamefont {Lo}\ \emph {et~al.}(2000)\citenamefont {Lo},
  \citenamefont {Wang}, \citenamefont {Dembo},\ and\ \citenamefont
  {Wang}}]{Lo:2000}%
  \BibitemOpen
  \bibfield  {author} {\bibinfo {author} {\bibfnamefont {C.-M.}\ \bibnamefont
  {Lo}}, \bibinfo {author} {\bibfnamefont {H.-B.}\ \bibnamefont {Wang}},
  \bibinfo {author} {\bibfnamefont {M.}~\bibnamefont {Dembo}}, \ and\ \bibinfo
  {author} {\bibfnamefont {Y.-L.}\ \bibnamefont {Wang}},\ }\bibfield  {title}
  {\enquote {\bibinfo {title} {Cell movement is guided by the rigidity of the
  substrate},}\ }\href@noop {} {\bibfield  {journal} {\bibinfo  {journal}
  {Biophysical Journal}\ }\textbf {\bibinfo {volume} {79}},\ \bibinfo {pages}
  {144--152} (\bibinfo {year} {2000})}\BibitemShut {NoStop}%
\bibitem [{\citenamefont {Engler}\ \emph {et~al.}(2006)\citenamefont {Engler},
  \citenamefont {Sen}, \citenamefont {Sweeney},\ and\ \citenamefont
  {Discher}}]{Engler:2006}%
  \BibitemOpen
  \bibfield  {author} {\bibinfo {author} {\bibfnamefont {A.}~\bibnamefont
  {Engler}}, \bibinfo {author} {\bibfnamefont {S.}~\bibnamefont {Sen}},
  \bibinfo {author} {\bibfnamefont {H.}~\bibnamefont {Sweeney}}, \ and\
  \bibinfo {author} {\bibfnamefont {D.}~\bibnamefont {Discher}},\ }\bibfield
  {title} {\enquote {\bibinfo {title} {Matrix elasticity directs stem cell
  lineage specification},}\ }\href@noop {} {\bibfield  {journal} {\bibinfo
  {journal} {Cell}\ }\textbf {\bibinfo {volume} {126}},\ \bibinfo {pages}
  {677--689} (\bibinfo {year} {2006})}\BibitemShut {NoStop}%
\bibitem [{\citenamefont {Miyoshi}, \citenamefont {Takaya},\ and\ \citenamefont
  {Nishinari}(1996)}]{Miyoshi:1996}%
  \BibitemOpen
  \bibfield  {author} {\bibinfo {author} {\bibfnamefont {E.}~\bibnamefont
  {Miyoshi}}, \bibinfo {author} {\bibfnamefont {T.}~\bibnamefont {Takaya}}, \
  and\ \bibinfo {author} {\bibfnamefont {K.}~\bibnamefont {Nishinari}},\
  }\bibfield  {title} {\enquote {\bibinfo {title} {Rheological and thermal
  studies of gel-sol transition in gellan gum aqueous solutions},}\ }\href@noop
  {} {\bibfield  {journal} {\bibinfo  {journal} {Carbohydrate Polymers}\
  }\textbf {\bibinfo {volume} {30}},\ \bibinfo {pages} {109--119} (\bibinfo
  {year} {1996})}\BibitemShut {NoStop}%
\bibitem [{\citenamefont {Djabourov}, \citenamefont {Leblond},\ and\
  \citenamefont {Papon}(1988)}]{Djabourov:1988}%
  \BibitemOpen
  \bibfield  {author} {\bibinfo {author} {\bibfnamefont {M.}~\bibnamefont
  {Djabourov}}, \bibinfo {author} {\bibfnamefont {J.}~\bibnamefont {Leblond}},
  \ and\ \bibinfo {author} {\bibfnamefont {P.}~\bibnamefont {Papon}},\
  }\bibfield  {title} {\enquote {\bibinfo {title} {Gelation of aqueous gelatin
  solutions. ii. rheology of the sol-gel transition},}\ }\href@noop {}
  {\bibfield  {journal} {\bibinfo  {journal} {J. Phys. France}\ }\textbf
  {\bibinfo {volume} {49}},\ \bibinfo {pages} {333--343} (\bibinfo {year}
  {1988})}\BibitemShut {NoStop}%
\bibitem [{\citenamefont {Mohammed}\ \emph {et~al.}(1998)\citenamefont
  {Mohammed}, \citenamefont {Hember}, \citenamefont {Richardson},\ and\
  \citenamefont {Morris}}]{Mohammed:1998}%
  \BibitemOpen
  \bibfield  {author} {\bibinfo {author} {\bibfnamefont {Z.}~\bibnamefont
  {Mohammed}}, \bibinfo {author} {\bibfnamefont {M.}~\bibnamefont {Hember}},
  \bibinfo {author} {\bibfnamefont {R.}~\bibnamefont {Richardson}}, \ and\
  \bibinfo {author} {\bibfnamefont {E.}~\bibnamefont {Morris}},\ }\bibfield
  {title} {\enquote {\bibinfo {title} {Kinetic and equilibrium processes in the
  formation and melting of agarose gels},}\ }\href@noop {} {\bibfield
  {journal} {\bibinfo  {journal} {Carbohydrate Polymers}\ }\textbf {\bibinfo
  {volume} {36}},\ \bibinfo {pages} {15--26} (\bibinfo {year}
  {1998})}\BibitemShut {NoStop}%
\bibitem [{\citenamefont {Norziah}, \citenamefont {Foo},\ and\ \citenamefont
  {Karim}(2006)}]{Norziah:2006}%
  \BibitemOpen
  \bibfield  {author} {\bibinfo {author} {\bibfnamefont {M.}~\bibnamefont
  {Norziah}}, \bibinfo {author} {\bibfnamefont {S.}~\bibnamefont {Foo}}, \ and\
  \bibinfo {author} {\bibfnamefont {A.}~\bibnamefont {Karim}},\ }\bibfield
  {title} {\enquote {\bibinfo {title} {Rheological studies on mixtures of agar
  (gracilaria changii) and k-carrageenan},}\ }\href@noop {} {\bibfield
  {journal} {\bibinfo  {journal} {Food Hydrocolloids}\ }\textbf {\bibinfo
  {volume} {20}},\ \bibinfo {pages} {206--217} (\bibinfo {year}
  {2006})}\BibitemShut {NoStop}%
\bibitem [{\citenamefont {Nordqvist}\ and\ \citenamefont
  {Vilgis}(2011)}]{Nordqvist:2011}%
  \BibitemOpen
  \bibfield  {author} {\bibinfo {author} {\bibfnamefont {D.}~\bibnamefont
  {Nordqvist}}\ and\ \bibinfo {author} {\bibfnamefont {T.}~\bibnamefont
  {Vilgis}},\ }\bibfield  {title} {\enquote {\bibinfo {title} {Rheological
  study of the gelation process of agarose-based solutions},}\ }\href@noop {}
  {\bibfield  {journal} {\bibinfo  {journal} {Food Biophysics}\ }\textbf
  {\bibinfo {volume} {6}},\ \bibinfo {pages} {450--460} (\bibinfo {year}
  {2011})}\BibitemShut {NoStop}%
\bibitem [{\citenamefont {Maurer}, \citenamefont {Junghans},\ and\
  \citenamefont {Vilgis}(2012)}]{Maurer:2012}%
  \BibitemOpen
  \bibfield  {author} {\bibinfo {author} {\bibfnamefont {S.}~\bibnamefont
  {Maurer}}, \bibinfo {author} {\bibfnamefont {A.}~\bibnamefont {Junghans}}, \
  and\ \bibinfo {author} {\bibfnamefont {T.}~\bibnamefont {Vilgis}},\
  }\bibfield  {title} {\enquote {\bibinfo {title} {Impact of xanthan gum,
  sucrose and fructose on the viscoelastic properties of agarose hydrogels},}\
  }\href@noop {} {\bibfield  {journal} {\bibinfo  {journal} {Food
  Hydrocolloids}\ }\textbf {\bibinfo {volume} {29}},\ \bibinfo {pages}
  {298--307} (\bibinfo {year} {2012})}\BibitemShut {NoStop}%
\bibitem [{\citenamefont {Richardson}\ and\ \citenamefont
  {Goycoolea}(1994)}]{Richardson:1994}%
  \BibitemOpen
  \bibfield  {author} {\bibinfo {author} {\bibfnamefont {R.}~\bibnamefont
  {Richardson}}\ and\ \bibinfo {author} {\bibfnamefont {F.}~\bibnamefont
  {Goycoolea}},\ }\bibfield  {title} {\enquote {\bibinfo {title} {Rheological
  measurement of $\kappa$-carrageenan during gelation},}\ }\href@noop {}
  {\bibfield  {journal} {\bibinfo  {journal} {Carbohydrate Polymers}\ }\textbf
  {\bibinfo {volume} {24}},\ \bibinfo {pages} {223--225} (\bibinfo {year}
  {1994})}\BibitemShut {NoStop}%
\bibitem [{\citenamefont {Normand}\ \emph {et~al.}(2000)\citenamefont
  {Normand}, \citenamefont {D.L.Lootens}, \citenamefont {Amici}, \citenamefont
  {Plucknett},\ and\ \citenamefont {Aymard}}]{Normand:2000}%
  \BibitemOpen
  \bibfield  {author} {\bibinfo {author} {\bibfnamefont {V.}~\bibnamefont
  {Normand}}, \bibinfo {author} {\bibnamefont {D.L.Lootens}}, \bibinfo {author}
  {\bibfnamefont {E.}~\bibnamefont {Amici}}, \bibinfo {author} {\bibfnamefont
  {K.}~\bibnamefont {Plucknett}}, \ and\ \bibinfo {author} {\bibfnamefont
  {P.}~\bibnamefont {Aymard}},\ }\bibfield  {title} {\enquote {\bibinfo {title}
  {New insight into agarose gel mechanical properties},}\ }\href@noop {}
  {\bibfield  {journal} {\bibinfo  {journal} {Biomacromolecules}\ }\textbf
  {\bibinfo {volume} {1}},\ \bibinfo {pages} {730--738} (\bibinfo {year}
  {2000})}\BibitemShut {NoStop}%
\bibitem [{\citenamefont {Ikeda}\ and\ \citenamefont
  {Nishinari}(2001)}]{Ikeda:2001}%
  \BibitemOpen
  \bibfield  {author} {\bibinfo {author} {\bibfnamefont {S.}~\bibnamefont
  {Ikeda}}\ and\ \bibinfo {author} {\bibfnamefont {K.}~\bibnamefont
  {Nishinari}},\ }\bibfield  {title} {\enquote {\bibinfo {title} {``weak
  gel"-type rheological properties of aqueous dispersions of nonaggregated
  $\kappa$-carrageenan helices},}\ }\href@noop {} {\bibfield  {journal}
  {\bibinfo  {journal} {J. Agric. Food. Chem.}\ }\textbf {\bibinfo {volume}
  {49}},\ \bibinfo {pages} {4436--4441} (\bibinfo {year} {2001})}\BibitemShut
  {NoStop}%
\bibitem [{\citenamefont {Ewoldt}, \citenamefont {Johnston},\ and\
  \citenamefont {Caretta}(2015)}]{Ewoldt:2015}%
  \BibitemOpen
  \bibfield  {author} {\bibinfo {author} {\bibfnamefont {R.}~\bibnamefont
  {Ewoldt}}, \bibinfo {author} {\bibfnamefont {M.}~\bibnamefont {Johnston}}, \
  and\ \bibinfo {author} {\bibfnamefont {L.}~\bibnamefont {Caretta}},\
  }\enquote {\bibinfo {title} {Complex fluids in biological systems},}\ \
  (\bibinfo  {publisher} {Springer New York},\ \bibinfo {year} {2015})\ Chap.\
  \bibinfo {chapter} {Chapter 6: Experimental Challenges of Shear Rheology: How
  to Avoid Bad Data}, pp.\ \bibinfo {pages} {207--241}\BibitemShut {NoStop}%
\bibitem [{\citenamefont {Mills}\ and\ \citenamefont
  {Snabre}(2009)}]{Mills:2009}%
  \BibitemOpen
  \bibfield  {author} {\bibinfo {author} {\bibfnamefont {P.}~\bibnamefont
  {Mills}}\ and\ \bibinfo {author} {\bibfnamefont {P.}~\bibnamefont {Snabre}},\
  }\bibfield  {title} {\enquote {\bibinfo {title} {Apparent viscosity and
  particle pressure of a concentrated suspension of non-brownian hard spheres
  near the jamming transition},}\ }\href@noop {} {\bibfield  {journal}
  {\bibinfo  {journal} {Eur. Phys. J. E}\ }\textbf {\bibinfo {volume} {30}},\
  \bibinfo {pages} {309--316} (\bibinfo {year} {2009})}\BibitemShut {NoStop}%
\bibitem [{\citenamefont {Boyer}, \citenamefont {Guazzelli},\ and\
  \citenamefont {Pouliquen}(2011)}]{Boyer:2011}%
  \BibitemOpen
  \bibfield  {author} {\bibinfo {author} {\bibfnamefont {F.}~\bibnamefont
  {Boyer}}, \bibinfo {author} {\bibfnamefont {E.}~\bibnamefont {Guazzelli}}, \
  and\ \bibinfo {author} {\bibfnamefont {O.}~\bibnamefont {Pouliquen}},\
  }\bibfield  {title} {\enquote {\bibinfo {title} {Unifying suspension and
  granular rheology},}\ }\href@noop {} {\bibfield  {journal} {\bibinfo
  {journal} {Phys. Rev. Lett.}\ }\textbf {\bibinfo {volume} {107}},\ \bibinfo
  {pages} {188301} (\bibinfo {year} {2011})}\BibitemShut {NoStop}%
\bibitem [{\citenamefont {Fall}\ \emph {et~al.}(2015)\citenamefont {Fall},
  \citenamefont {Ovarlez}, \citenamefont {Hautemayou}, \citenamefont
  {M\'ezi\`ere}, \citenamefont {Roux},\ and\ \citenamefont
  {Chevoir}}]{Fall:2015}%
  \BibitemOpen
  \bibfield  {author} {\bibinfo {author} {\bibfnamefont {A.}~\bibnamefont
  {Fall}}, \bibinfo {author} {\bibfnamefont {G.}~\bibnamefont {Ovarlez}},
  \bibinfo {author} {\bibfnamefont {D.}~\bibnamefont {Hautemayou}}, \bibinfo
  {author} {\bibfnamefont {C.}~\bibnamefont {M\'ezi\`ere}}, \bibinfo {author}
  {\bibfnamefont {J.-N.}\ \bibnamefont {Roux}}, \ and\ \bibinfo {author}
  {\bibfnamefont {F.}~\bibnamefont {Chevoir}},\ }\bibfield  {title} {\enquote
  {\bibinfo {title} {Dry granular flows: Rheological measurements of the
  $\mu(i)$-rheology},}\ }\href@noop {} {\bibfield  {journal} {\bibinfo
  {journal} {J. Rheol.}\ }\textbf {\bibinfo {volume} {59}},\ \bibinfo {pages}
  {1065--1080} (\bibinfo {year} {2015})}\BibitemShut {NoStop}%
\bibitem [{\citenamefont {Mezger}(2015)}]{Flow:2015}%
  \BibitemOpen
  \bibfield  {author} {\bibinfo {author} {\bibfnamefont {T.}~\bibnamefont
  {Mezger}},\ }\href@noop {} {\enquote {\bibinfo {title} {Tips and tricks from
  joe flow - normal force control: May the force be with you},}\ }\bibinfo
  {type} {Tech. Rep.}\ (\bibinfo  {institution} {Anton Paar},\ \bibinfo {year}
  {2015})\BibitemShut {NoStop}%
\bibitem [{\citenamefont {Araki}(1956)}]{Araki:1956}%
  \BibitemOpen
  \bibfield  {author} {\bibinfo {author} {\bibfnamefont {C.}~\bibnamefont
  {Araki}},\ }\bibfield  {title} {\enquote {\bibinfo {title} {Structure of the
  agarose consitutent of agar-agar},}\ }\href@noop {} {\bibfield  {journal}
  {\bibinfo  {journal} {Bulletin of the Chemical Society of Japan}\ }\textbf
  {\bibinfo {volume} {29}},\ \bibinfo {pages} {543--544} (\bibinfo {year}
  {1956})}\BibitemShut {NoStop}%
\bibitem [{\citenamefont {Stanley}(2006)}]{Stanley:2006}%
  \BibitemOpen
  \bibfield  {author} {\bibinfo {author} {\bibfnamefont {N.}~\bibnamefont
  {Stanley}},\ }\enquote {\bibinfo {title} {Food polysaccharides and their
  applications, second edition},}\ \ (\bibinfo  {publisher} {A.M. stephens and
  G.O. Philips and P.A. Williams},\ \bibinfo {year} {2006})\ Chap.\ \bibinfo
  {chapter} {agars}, pp.\ \bibinfo {pages} {217--238}\BibitemShut {NoStop}%
\bibitem [{\citenamefont {Feke}\ and\ \citenamefont {Prins}(1974)}]{Feke:1974}%
  \BibitemOpen
  \bibfield  {author} {\bibinfo {author} {\bibfnamefont {G.}~\bibnamefont
  {Feke}}\ and\ \bibinfo {author} {\bibfnamefont {W.}~\bibnamefont {Prins}},\
  }\bibfield  {title} {\enquote {\bibinfo {title} {Spinodal phase separation in
  a macromolecular sol-gel transition},}\ }\href@noop {} {\bibfield  {journal}
  {\bibinfo  {journal} {Macromolecules}\ }\textbf {\bibinfo {volume} {7}},\
  \bibinfo {pages} {527--530} (\bibinfo {year} {1974})}\BibitemShut {NoStop}%
\bibitem [{\citenamefont {SanBiagio}\ \emph {et~al.}(1996)\citenamefont
  {SanBiagio}, \citenamefont {Bulone}, \citenamefont {Emanuele}, \citenamefont
  {Palma-Vittorelli},\ and\ \citenamefont {Palma}}]{SanBiagio:1996}%
  \BibitemOpen
  \bibfield  {author} {\bibinfo {author} {\bibfnamefont {P.}~\bibnamefont
  {SanBiagio}}, \bibinfo {author} {\bibfnamefont {D.}~\bibnamefont {Bulone}},
  \bibinfo {author} {\bibfnamefont {A.}~\bibnamefont {Emanuele}}, \bibinfo
  {author} {\bibfnamefont {M.}~\bibnamefont {Palma-Vittorelli}}, \ and\
  \bibinfo {author} {\bibfnamefont {M.}~\bibnamefont {Palma}},\ }\bibfield
  {title} {\enquote {\bibinfo {title} {Spontaneous symmetry-breaking pathways:
  time-resolved study of agarose gelation},}\ }\href@noop {} {\bibfield
  {journal} {\bibinfo  {journal} {Food Hydrocolloids}\ }\textbf {\bibinfo
  {volume} {10}},\ \bibinfo {pages} {91--97} (\bibinfo {year}
  {1996})}\BibitemShut {NoStop}%
\bibitem [{\citenamefont {Matsuo}, \citenamefont {Tanaka},\ and\ \citenamefont
  {Ma}(2002)}]{Matsuo:2002}%
  \BibitemOpen
  \bibfield  {author} {\bibinfo {author} {\bibfnamefont {M.}~\bibnamefont
  {Matsuo}}, \bibinfo {author} {\bibfnamefont {T.}~\bibnamefont {Tanaka}}, \
  and\ \bibinfo {author} {\bibfnamefont {L.}~\bibnamefont {Ma}},\ }\bibfield
  {title} {\enquote {\bibinfo {title} {Gelation mechanism of agarose and
  $\kappa$-carrageenan solutions estimated in terms of concentration
  fluctuation},}\ }\href@noop {} {\bibfield  {journal} {\bibinfo  {journal}
  {Polymer}\ }\textbf {\bibinfo {volume} {43}},\ \bibinfo {pages} {5299--5309}
  (\bibinfo {year} {2002})}\BibitemShut {NoStop}%
\bibitem [{\citenamefont {Tako}\ and\ \citenamefont
  {Nakamura}(1988)}]{Tako:1988}%
  \BibitemOpen
  \bibfield  {author} {\bibinfo {author} {\bibfnamefont {M.}~\bibnamefont
  {Tako}}\ and\ \bibinfo {author} {\bibfnamefont {S.}~\bibnamefont
  {Nakamura}},\ }\bibfield  {title} {\enquote {\bibinfo {title} {Gelation
  mechanism of agarose},}\ }\href@noop {} {\bibfield  {journal} {\bibinfo
  {journal} {Carbohydrate Research}\ }\textbf {\bibinfo {volume} {180}},\
  \bibinfo {pages} {277--284} (\bibinfo {year} {1988})}\BibitemShut {NoStop}%
\bibitem [{\citenamefont {Braudo}(1992)}]{Braudo:1992}%
  \BibitemOpen
  \bibfield  {author} {\bibinfo {author} {\bibfnamefont {E.}~\bibnamefont
  {Braudo}},\ }\bibfield  {title} {\enquote {\bibinfo {title} {Mechanism for
  galactan gelation},}\ }\href@noop {} {\bibfield  {journal} {\bibinfo
  {journal} {Food Hydrocolloids}\ }\textbf {\bibinfo {volume} {6}},\ \bibinfo
  {pages} {25--43} (\bibinfo {year} {1992})}\BibitemShut {NoStop}%
\bibitem [{\citenamefont {Manno}\ \emph {et~al.}(1999)\citenamefont {Manno},
  \citenamefont {Emanuele}, \citenamefont {Martorana}, \citenamefont {Bulone},
  \citenamefont {Biagio}, \citenamefont {Palma-Vittorelli},\ and\ \citenamefont
  {Palma}}]{Manno:1999}%
  \BibitemOpen
  \bibfield  {author} {\bibinfo {author} {\bibfnamefont {M.}~\bibnamefont
  {Manno}}, \bibinfo {author} {\bibfnamefont {E.}~\bibnamefont {Emanuele}},
  \bibinfo {author} {\bibfnamefont {V.}~\bibnamefont {Martorana}}, \bibinfo
  {author} {\bibfnamefont {D.}~\bibnamefont {Bulone}}, \bibinfo {author}
  {\bibfnamefont {P.~S.}\ \bibnamefont {Biagio}}, \bibinfo {author}
  {\bibfnamefont {M.}~\bibnamefont {Palma-Vittorelli}}, \ and\ \bibinfo
  {author} {\bibfnamefont {M.}~\bibnamefont {Palma}},\ }\bibfield  {title}
  {\enquote {\bibinfo {title} {Multiple interactions between molecular and
  supramolecular ordering},}\ }\href@noop {} {\bibfield  {journal} {\bibinfo
  {journal} {Phys. Rev. E}\ }\textbf {\bibinfo {volume} {59}},\ \bibinfo
  {pages} {2222--2230} (\bibinfo {year} {1999})}\BibitemShut {NoStop}%
\bibitem [{\citenamefont {Aymard}\ \emph {et~al.}(2001)\citenamefont {Aymard},
  \citenamefont {Martin}, \citenamefont {Plucknett}, \citenamefont {Foster},
  \citenamefont {Clark},\ and\ \citenamefont {Norton}}]{Aymard:2001}%
  \BibitemOpen
  \bibfield  {author} {\bibinfo {author} {\bibfnamefont {P.}~\bibnamefont
  {Aymard}}, \bibinfo {author} {\bibfnamefont {D.}~\bibnamefont {Martin}},
  \bibinfo {author} {\bibfnamefont {K.}~\bibnamefont {Plucknett}}, \bibinfo
  {author} {\bibfnamefont {T.}~\bibnamefont {Foster}}, \bibinfo {author}
  {\bibfnamefont {A.}~\bibnamefont {Clark}}, \ and\ \bibinfo {author}
  {\bibfnamefont {I.}~\bibnamefont {Norton}},\ }\bibfield  {title} {\enquote
  {\bibinfo {title} {Influence of thermal history on the structural and
  mechanical properties of agarose gels},}\ }\href@noop {} {\bibfield
  {journal} {\bibinfo  {journal} {Biopolymers}\ }\textbf {\bibinfo {volume}
  {59}},\ \bibinfo {pages} {131--144} (\bibinfo {year} {2001})}\BibitemShut
  {NoStop}%
\bibitem [{\citenamefont {Xiong}\ \emph {et~al.}(2005)\citenamefont {Xiong},
  \citenamefont {Narayanan}, \citenamefont {Liu}, \citenamefont {Chong},
  \citenamefont {Chen},\ and\ \citenamefont {Chung}}]{Xiong:2005}%
  \BibitemOpen
  \bibfield  {author} {\bibinfo {author} {\bibfnamefont {J.-Y.}\ \bibnamefont
  {Xiong}}, \bibinfo {author} {\bibfnamefont {J.}~\bibnamefont {Narayanan}},
  \bibinfo {author} {\bibfnamefont {X.-Y.}\ \bibnamefont {Liu}}, \bibinfo
  {author} {\bibfnamefont {T.}~\bibnamefont {Chong}}, \bibinfo {author}
  {\bibfnamefont {S.}~\bibnamefont {Chen}}, \ and\ \bibinfo {author}
  {\bibfnamefont {T.-S.}\ \bibnamefont {Chung}},\ }\bibfield  {title} {\enquote
  {\bibinfo {title} {Topology evolution and gelation mechanism of agarose
  gel},}\ }\href@noop {} {\bibfield  {journal} {\bibinfo  {journal} {J. Phys.
  Chem. B}\ }\textbf {\bibinfo {volume} {109}},\ \bibinfo {pages} {5638--5643}
  (\bibinfo {year} {2005})}\BibitemShut {NoStop}%
\bibitem [{\citenamefont {Arnott}, \citenamefont {Fulmer},\ and\ \citenamefont
  {Scott}(1974)}]{Arnott:1974}%
  \BibitemOpen
  \bibfield  {author} {\bibinfo {author} {\bibfnamefont {S.}~\bibnamefont
  {Arnott}}, \bibinfo {author} {\bibfnamefont {A.}~\bibnamefont {Fulmer}}, \
  and\ \bibinfo {author} {\bibfnamefont {W.}~\bibnamefont {Scott}},\ }\bibfield
   {title} {\enquote {\bibinfo {title} {The agarose double helix and its
  function in agarose gel structure},}\ }\href@noop {} {\bibfield  {journal}
  {\bibinfo  {journal} {Journal of Molecular Biology}\ }\textbf {\bibinfo
  {volume} {90}},\ \bibinfo {pages} {269--284} (\bibinfo {year}
  {1974})}\BibitemShut {NoStop}%
\bibitem [{\citenamefont {Clark}\ and\ \citenamefont
  {Ross-Murphy}(1987)}]{Clark:1987}%
  \BibitemOpen
  \bibfield  {author} {\bibinfo {author} {\bibfnamefont {A.}~\bibnamefont
  {Clark}}\ and\ \bibinfo {author} {\bibfnamefont {S.}~\bibnamefont
  {Ross-Murphy}},\ }\bibfield  {title} {\enquote {\bibinfo {title} {Structural
  and mechanical properties of biopolymer gels},}\ }\href@noop {} {\bibfield
  {journal} {\bibinfo  {journal} {Advances in Polymer Science}\ }\textbf
  {\bibinfo {volume} {83}},\ \bibinfo {pages} {57--192} (\bibinfo {year}
  {1987})}\BibitemShut {NoStop}%
\bibitem [{\citenamefont {Foord}, \citenamefont {Atkins},\ and\ \citenamefont
  {Wills}(1989)}]{Foord:1989}%
  \BibitemOpen
  \bibfield  {author} {\bibinfo {author} {\bibfnamefont {S.}~\bibnamefont
  {Foord}}, \bibinfo {author} {\bibfnamefont {E.}~\bibnamefont {Atkins}}, \
  and\ \bibinfo {author} {\bibfnamefont {H.}~\bibnamefont {Wills}},\ }\bibfield
   {title} {\enquote {\bibinfo {title} {New x-ray diffraction results from
  agarose: Extended single helix structures and implications for gelation
  mechanism},}\ }\href@noop {} {\bibfield  {journal} {\bibinfo  {journal}
  {Biopolymers}\ }\textbf {\bibinfo {volume} {28}},\ \bibinfo {pages}
  {1345--1365} (\bibinfo {year} {1989})}\BibitemShut {NoStop}%
\bibitem [{\citenamefont {Djabourov}\ \emph {et~al.}(1989)\citenamefont
  {Djabourov}, \citenamefont {Clark}, \citenamefont {Rowlands},\ and\
  \citenamefont {Ross-Murphy}}]{Djabourov:1989}%
  \BibitemOpen
  \bibfield  {author} {\bibinfo {author} {\bibfnamefont {M.}~\bibnamefont
  {Djabourov}}, \bibinfo {author} {\bibfnamefont {A.}~\bibnamefont {Clark}},
  \bibinfo {author} {\bibfnamefont {D.}~\bibnamefont {Rowlands}}, \ and\
  \bibinfo {author} {\bibfnamefont {S.}~\bibnamefont {Ross-Murphy}},\
  }\bibfield  {title} {\enquote {\bibinfo {title} {Small-angle x-ray scattering
  characterization of agarose sols and gels},}\ }\href@noop {} {\bibfield
  {journal} {\bibinfo  {journal} {Macromolecules}\ }\textbf {\bibinfo {volume}
  {22}},\ \bibinfo {pages} {160--180} (\bibinfo {year} {1989})}\BibitemShut
  {NoStop}%
\bibitem [{\citenamefont {Schafer}\ and\ \citenamefont
  {Stevens}(1995)}]{Schafer:1995}%
  \BibitemOpen
  \bibfield  {author} {\bibinfo {author} {\bibfnamefont {S.}~\bibnamefont
  {Schafer}}\ and\ \bibinfo {author} {\bibfnamefont {E.}~\bibnamefont
  {Stevens}},\ }\bibfield  {title} {\enquote {\bibinfo {title} {A reexamination
  of the double-helix model for agarose gels using optical rotation},}\
  }\href@noop {} {\bibfield  {journal} {\bibinfo  {journal} {Biopolymers}\
  }\textbf {\bibinfo {volume} {36}},\ \bibinfo {pages} {103--108} (\bibinfo
  {year} {1995})}\BibitemShut {NoStop}%
\bibitem [{\citenamefont {Guenet}\ and\ \citenamefont
  {Rochas}(2006)}]{Guenet:2006}%
  \BibitemOpen
  \bibfield  {author} {\bibinfo {author} {\bibfnamefont {J.-M.}\ \bibnamefont
  {Guenet}}\ and\ \bibinfo {author} {\bibfnamefont {C.}~\bibnamefont
  {Rochas}},\ }\bibfield  {title} {\enquote {\bibinfo {title} {Agarose sols and
  gels revisited},}\ }\href@noop {} {\bibfield  {journal} {\bibinfo  {journal}
  {Macromol. Symp.}\ }\textbf {\bibinfo {volume} {242}},\ \bibinfo {pages}
  {65--70} (\bibinfo {year} {2006})}\BibitemShut {NoStop}%
\bibitem [{\citenamefont {Chui}, \citenamefont {Philips},\ and\ \citenamefont
  {McCarthy}(1995)}]{Chui:1995}%
  \BibitemOpen
  \bibfield  {author} {\bibinfo {author} {\bibfnamefont {M.}~\bibnamefont
  {Chui}}, \bibinfo {author} {\bibfnamefont {R.}~\bibnamefont {Philips}}, \
  and\ \bibinfo {author} {\bibfnamefont {M.}~\bibnamefont {McCarthy}},\
  }\bibfield  {title} {\enquote {\bibinfo {title} {Measurement of the porous
  microstructure of hydrogels by nuclear magnetic resonance},}\ }\href@noop {}
  {\bibfield  {journal} {\bibinfo  {journal} {Journal of Colloid and Interface
  Science}\ }\textbf {\bibinfo {volume} {174}},\ \bibinfo {pages} {336--344}
  (\bibinfo {year} {1995})}\BibitemShut {NoStop}%
\bibitem [{\citenamefont {Pernodet}, \citenamefont {Maaloum},\ and\
  \citenamefont {Tinland}(1997)}]{Pernodet:1997}%
  \BibitemOpen
  \bibfield  {author} {\bibinfo {author} {\bibfnamefont {N.}~\bibnamefont
  {Pernodet}}, \bibinfo {author} {\bibfnamefont {M.}~\bibnamefont {Maaloum}}, \
  and\ \bibinfo {author} {\bibfnamefont {B.}~\bibnamefont {Tinland}},\
  }\bibfield  {title} {\enquote {\bibinfo {title} {Pore sizes of agarose gels
  by atomic force microscopy},}\ }\href@noop {} {\bibfield  {journal} {\bibinfo
   {journal} {Electrophoresis}\ }\textbf {\bibinfo {volume} {18}},\ \bibinfo
  {pages} {55--58} (\bibinfo {year} {1997})}\BibitemShut {NoStop}%
\bibitem [{\citenamefont {Bonn}\ \emph {et~al.}(1998)\citenamefont {Bonn},
  \citenamefont {Kellay}, \citenamefont {Prochnow}, \citenamefont
  {Ben-Djemiaa},\ and\ \citenamefont {Meunier}}]{Bonn:1998b}%
  \BibitemOpen
  \bibfield  {author} {\bibinfo {author} {\bibfnamefont {D.}~\bibnamefont
  {Bonn}}, \bibinfo {author} {\bibfnamefont {H.}~\bibnamefont {Kellay}},
  \bibinfo {author} {\bibfnamefont {M.}~\bibnamefont {Prochnow}}, \bibinfo
  {author} {\bibfnamefont {K.}~\bibnamefont {Ben-Djemiaa}}, \ and\ \bibinfo
  {author} {\bibfnamefont {J.}~\bibnamefont {Meunier}},\ }\bibfield  {title}
  {\enquote {\bibinfo {title} {Delayed fracture of an inhomogeneous soft
  solid},}\ }\href@noop {} {\bibfield  {journal} {\bibinfo  {journal}
  {Science}\ }\textbf {\bibinfo {volume} {280}},\ \bibinfo {pages} {265--267}
  (\bibinfo {year} {1998})}\BibitemShut {NoStop}%
\bibitem [{\citenamefont {Barrangou}, \citenamefont {Daubert},\ and\
  \citenamefont {Foegeding}(2006)}]{Barrangou:2006}%
  \BibitemOpen
  \bibfield  {author} {\bibinfo {author} {\bibfnamefont {L.}~\bibnamefont
  {Barrangou}}, \bibinfo {author} {\bibfnamefont {C.}~\bibnamefont {Daubert}},
  \ and\ \bibinfo {author} {\bibfnamefont {E.}~\bibnamefont {Foegeding}},\
  }\bibfield  {title} {\enquote {\bibinfo {title} {Textural properties of
  agarose gels. i. rheological and fracture properties},}\ }\href@noop {}
  {\bibfield  {journal} {\bibinfo  {journal} {Food Hydrocolloids}\ }\textbf
  {\bibinfo {volume} {20}},\ \bibinfo {pages} {184--195} (\bibinfo {year}
  {2006})}\BibitemShut {NoStop}%
\bibitem [{\citenamefont {Daniels}\ \emph {et~al.}(2007)\citenamefont
  {Daniels}, \citenamefont {Mukhopadhyay}, \citenamefont {Houseworth},\ and\
  \citenamefont {R.P.Behringer}}]{Daniels:2007}%
  \BibitemOpen
  \bibfield  {author} {\bibinfo {author} {\bibfnamefont {K.}~\bibnamefont
  {Daniels}}, \bibinfo {author} {\bibfnamefont {S.}~\bibnamefont
  {Mukhopadhyay}}, \bibinfo {author} {\bibfnamefont {P.}~\bibnamefont
  {Houseworth}}, \ and\ \bibinfo {author} {\bibnamefont {R.P.Behringer}},\
  }\bibfield  {title} {\enquote {\bibinfo {title} {Instabilities in droplets
  spreading on gels},}\ }\href@noop {} {\bibfield  {journal} {\bibinfo
  {journal} {Phys. Rev. Lett.}\ }\textbf {\bibinfo {volume} {99}},\ \bibinfo
  {pages} {124501} (\bibinfo {year} {2007})}\BibitemShut {NoStop}%
\bibitem [{\citenamefont {Spandagos}\ \emph {et~al.}(2012)\citenamefont
  {Spandagos}, \citenamefont {Goudoulas}, \citenamefont {Luckham},\ and\
  \citenamefont {Matar}}]{Spandagos:2012}%
  \BibitemOpen
  \bibfield  {author} {\bibinfo {author} {\bibfnamefont {C.}~\bibnamefont
  {Spandagos}}, \bibinfo {author} {\bibfnamefont {T.}~\bibnamefont
  {Goudoulas}}, \bibinfo {author} {\bibfnamefont {P.}~\bibnamefont {Luckham}},
  \ and\ \bibinfo {author} {\bibfnamefont {O.}~\bibnamefont {Matar}},\
  }\bibfield  {title} {\enquote {\bibinfo {title} {Surface tension-induced gel
  fracture. part 1. fracture of agar gels},}\ }\href@noop {} {\bibfield
  {journal} {\bibinfo  {journal} {Langmuir}\ }\textbf {\bibinfo {volume}
  {28}},\ \bibinfo {pages} {7197--7211} (\bibinfo {year} {2012})}\BibitemShut
  {NoStop}%
\bibitem [{\citenamefont {Matsuhashi}(1990)}]{Matsuhashi:1990}%
  \BibitemOpen
  \bibfield  {author} {\bibinfo {author} {\bibfnamefont {T.}~\bibnamefont
  {Matsuhashi}},\ }\enquote {\bibinfo {title} {Food gels},}\ \ (\bibinfo
  {publisher} {Elsevier Science Publishers LTD},\ \bibinfo {year} {1990})\
  Chap.\ \bibinfo {chapter} {1. Agar}, pp.\ \bibinfo {pages}
  {1--51}\BibitemShut {NoStop}%
\bibitem [{\citenamefont {Nakayama}\ \emph {et~al.}(1978)\citenamefont
  {Nakayama}, \citenamefont {Kawasaki}, \citenamefont {Niwa},\ and\
  \citenamefont {Hamada}}]{Nakayama:1978}%
  \BibitemOpen
  \bibfield  {author} {\bibinfo {author} {\bibfnamefont {T.}~\bibnamefont
  {Nakayama}}, \bibinfo {author} {\bibfnamefont {M.}~\bibnamefont {Kawasaki}},
  \bibinfo {author} {\bibfnamefont {E.}~\bibnamefont {Niwa}}, \ and\ \bibinfo
  {author} {\bibfnamefont {I.}~\bibnamefont {Hamada}},\ }\bibfield  {title}
  {\enquote {\bibinfo {title} {Compression creep behavior and syneresis water
  of agar-agar and actomyosin gels},}\ }\href@noop {} {\bibfield  {journal}
  {\bibinfo  {journal} {Journal of Food Science}\ }\textbf {\bibinfo {volume}
  {43}},\ \bibinfo {pages} {1430--1432} (\bibinfo {year} {1978})}\BibitemShut
  {NoStop}%
\bibitem [{\citenamefont {Hickson}\ and\ \citenamefont
  {Polson}(1968)}]{Hickson:1968}%
  \BibitemOpen
  \bibfield  {author} {\bibinfo {author} {\bibfnamefont {T.}~\bibnamefont
  {Hickson}}\ and\ \bibinfo {author} {\bibfnamefont {A.}~\bibnamefont
  {Polson}},\ }\bibfield  {title} {\enquote {\bibinfo {title} {Some physical
  characteristics of the agarose molecule},}\ }\href@noop {} {\bibfield
  {journal} {\bibinfo  {journal} {Biochemica et Biophysica Acta}\ }\textbf
  {\bibinfo {volume} {165}},\ \bibinfo {pages} {43--58} (\bibinfo {year}
  {1968})}\BibitemShut {NoStop}%
\bibitem [{\citenamefont {Mao}\ \emph {et~al.}(2016)\citenamefont {Mao},
  \citenamefont {Bentaleb}, \citenamefont {Louerat}, \citenamefont {Divoux},\
  and\ \citenamefont {Snabre}}]{Mao:2015}%
  \BibitemOpen
  \bibfield  {author} {\bibinfo {author} {\bibfnamefont {B.}~\bibnamefont
  {Mao}}, \bibinfo {author} {\bibfnamefont {A.}~\bibnamefont {Bentaleb}},
  \bibinfo {author} {\bibfnamefont {F.}~\bibnamefont {Louerat}}, \bibinfo
  {author} {\bibfnamefont {T.}~\bibnamefont {Divoux}}, \ and\ \bibinfo {author}
  {\bibfnamefont {P.}~\bibnamefont {Snabre}},\ }\bibfield  {title} {\enquote
  {\bibinfo {title} {Overcooked agar solutions: impact on the structural and
  mechanical properties of agar gels},}\ }\href@noop {} {\bibfield  {journal}
  {\bibinfo  {journal} {Submitted to Soft Matter}\ } (\bibinfo {year} {February
  2016})}\BibitemShut {NoStop}%
\bibitem [{\citenamefont {Goycoolea}\ \emph {et~al.}(1995)\citenamefont
  {Goycoolea}, \citenamefont {Richardson}, \citenamefont {Morris},\ and\
  \citenamefont {Gidley}}]{Goycoolea:1995}%
  \BibitemOpen
  \bibfield  {author} {\bibinfo {author} {\bibfnamefont {F.}~\bibnamefont
  {Goycoolea}}, \bibinfo {author} {\bibfnamefont {R.}~\bibnamefont
  {Richardson}}, \bibinfo {author} {\bibfnamefont {E.}~\bibnamefont {Morris}},
  \ and\ \bibinfo {author} {\bibfnamefont {M.}~\bibnamefont {Gidley}},\
  }\bibfield  {title} {\enquote {\bibinfo {title} {Effect of locus bean gum and
  konjac glucomannan on the conformation and rheology of agarose and
  $\kappa$-carrageenan},}\ }\href@noop {} {\bibfield  {journal} {\bibinfo
  {journal} {Biopolymers}\ }\textbf {\bibinfo {volume} {36}},\ \bibinfo {pages}
  {643--658} (\bibinfo {year} {1995})}\BibitemShut {NoStop}%
\bibitem [{\citenamefont {Labropoulos}\ \emph {et~al.}(2002)\citenamefont
  {Labropoulos}, \citenamefont {Niesz}, \citenamefont {Danforth},\ and\
  \citenamefont {Kevrekidis}}]{Labropoulos:2002}%
  \BibitemOpen
  \bibfield  {author} {\bibinfo {author} {\bibfnamefont {K.}~\bibnamefont
  {Labropoulos}}, \bibinfo {author} {\bibfnamefont {D.}~\bibnamefont {Niesz}},
  \bibinfo {author} {\bibfnamefont {S.}~\bibnamefont {Danforth}}, \ and\
  \bibinfo {author} {\bibfnamefont {P.}~\bibnamefont {Kevrekidis}},\ }\bibfield
   {title} {\enquote {\bibinfo {title} {Dynamics rheology of agar gels: theory
  and experiments. part ii: gelation behavior of agar sols and fitting of a
  theoretical rheological model},}\ }\href@noop {} {\bibfield  {journal}
  {\bibinfo  {journal} {Carbohydrate Polymers}\ }\textbf {\bibinfo {volume}
  {50}},\ \bibinfo {pages} {407--415} (\bibinfo {year} {2002})}\BibitemShut
  {NoStop}%
\bibitem [{\citenamefont {Normand}\ \emph {et~al.}(2003)\citenamefont
  {Normand}, \citenamefont {Aymard}, \citenamefont {Lootens}, \citenamefont
  {Amici}, \citenamefont {Plucknett},\ and\ \citenamefont
  {Frith}}]{Normand:2003}%
  \BibitemOpen
  \bibfield  {author} {\bibinfo {author} {\bibfnamefont {V.}~\bibnamefont
  {Normand}}, \bibinfo {author} {\bibfnamefont {P.}~\bibnamefont {Aymard}},
  \bibinfo {author} {\bibfnamefont {D.}~\bibnamefont {Lootens}}, \bibinfo
  {author} {\bibfnamefont {E.}~\bibnamefont {Amici}}, \bibinfo {author}
  {\bibfnamefont {K.}~\bibnamefont {Plucknett}}, \ and\ \bibinfo {author}
  {\bibfnamefont {W.}~\bibnamefont {Frith}},\ }\bibfield  {title} {\enquote
  {\bibinfo {title} {Effect of sucrose on agarose gels mechanical behaviour},}\
  }\href@noop {} {\bibfield  {journal} {\bibinfo  {journal} {Carbohydrate
  Polymers}\ }\textbf {\bibinfo {volume} {54}},\ \bibinfo {pages} {83--95}
  (\bibinfo {year} {2003})}\BibitemShut {NoStop}%
\bibitem [{\citenamefont {Piazza}\ and\ \citenamefont
  {Benedetti}(2010)}]{Piazza:2010}%
  \BibitemOpen
  \bibfield  {author} {\bibinfo {author} {\bibfnamefont {L.}~\bibnamefont
  {Piazza}}\ and\ \bibinfo {author} {\bibfnamefont {S.}~\bibnamefont
  {Benedetti}},\ }\bibfield  {title} {\enquote {\bibinfo {title} {Investigation
  on the rheological properties of agar gels and their role on aroma release in
  agar/limonene solid emulsions},}\ }\href@noop {} {\bibfield  {journal}
  {\bibinfo  {journal} {Food Research International}\ }\textbf {\bibinfo
  {volume} {43}},\ \bibinfo {pages} {269--276} (\bibinfo {year}
  {2010})}\BibitemShut {NoStop}%
\bibitem [{Note1()}]{Note1}%
  \BibitemOpen
  \bibinfo {note} {Note that repeating the gelation experiment under zero
  normal force with other initial gap values, i.e. $e_0=200$~$\mu $m and
  1000~$\mu $m gives compatible values of $G'_f$ within error bars, which
  demonstrates that the value of $G'_f$ determined with the zero normal force
  protocol is independent of the initial value $e_0$ of the gap
  width.}\BibitemShut {Stop}%
\bibitem [{\citenamefont {Altmann}\ \emph {et~al.}(2004)\citenamefont
  {Altmann}, \citenamefont {Cooper-White}, \citenamefont {Dunstan},\ and\
  \citenamefont {Stokes}}]{Altmann:2004}%
  \BibitemOpen
  \bibfield  {author} {\bibinfo {author} {\bibfnamefont {N.}~\bibnamefont
  {Altmann}}, \bibinfo {author} {\bibfnamefont {J.}~\bibnamefont
  {Cooper-White}}, \bibinfo {author} {\bibfnamefont {D.}~\bibnamefont
  {Dunstan}}, \ and\ \bibinfo {author} {\bibfnamefont {J.}~\bibnamefont
  {Stokes}},\ }\bibfield  {title} {\enquote {\bibinfo {title} {Strong through
  to weak 'sheared' gels},}\ }\href@noop {} {\bibfield  {journal} {\bibinfo
   {journal} {J. Non-Newtonian Fluid Mech.}\ }\textbf {\bibinfo {volume}
  {124}},\ \bibinfo {pages} {129--136} (\bibinfo {year} {2004})}\BibitemShut
  {NoStop}%
\bibitem [{\citenamefont {Russ}, \citenamefont {Zielbauer},\ and\ \citenamefont
  {Vilgis}(2014)}]{Russ:2014}%
  \BibitemOpen
  \bibfield  {author} {\bibinfo {author} {\bibfnamefont {N.}~\bibnamefont
  {Russ}}, \bibinfo {author} {\bibfnamefont {B.}~\bibnamefont {Zielbauer}}, \
  and\ \bibinfo {author} {\bibfnamefont {T.}~\bibnamefont {Vilgis}},\
  }\bibfield  {title} {\enquote {\bibinfo {title} {Impact of sucrose and
  trehalose on different agarose-hydrocolloid systems},}\ }\href@noop {}
  {\bibfield  {journal} {\bibinfo  {journal} {Food Hydrocolloids}\ }\textbf
  {\bibinfo {volume} {41}},\ \bibinfo {pages} {44--52} (\bibinfo {year}
  {2014})}\BibitemShut {NoStop}%
\bibitem [{Note2()}]{Note2}%
  \BibitemOpen
  \bibinfo {note} {Note that a proper definition of the gelation point is the
  instant where $G'$ and $G''$ both scale as identical power laws of frequency
  which corresponds to the value of the phase angle $\delta = \protect \qopname
  \relax o{arctan}(G''/G')$ that is independent of the frequency \cite
  {Chambon:1987}. However, our goal here is not to determine a gelation point,
  but only to improve the determination of the intersection of $G'$ and
  $G''$.}\BibitemShut {Stop}%
\bibitem [{\citenamefont {Orafidiya}(1989)}]{Orafidiya:1989}%
  \BibitemOpen
  \bibfield  {author} {\bibinfo {author} {\bibfnamefont {L.}~\bibnamefont
  {Orafidiya}},\ }\bibfield  {title} {\enquote {\bibinfo {title} {Continuous
  shear rheometry of o/w emulsions; control of evaporation in cone/plate
  geometry},}\ }\href@noop {} {\bibfield  {journal} {\bibinfo  {journal} {J.
  Pharm. Pharmacol.}\ }\textbf {\bibinfo {volume} {41}},\ \bibinfo {pages}
  {341--342} (\bibinfo {year} {1989})}\BibitemShut {NoStop}%
\bibitem [{Note3()}]{Note3}%
  \BibitemOpen
  \bibinfo {note} {We checked that the surrounding oil does not impact the
  steady state value of the elastic modulus $G'_f$ by performing an experiment
  over a much shorter duration. Two experiments conducted at a cooling rate
  $\protect \mathaccentV {dot}05F{\protect \rm T}=1^{\circ }$C/min, one with
  water in the solvent trap and the other one with an oil layer around the
  sample lead to compatible values of the elastic modulus, within error
  bars.}\BibitemShut {Stop}%
\bibitem [{Note4()}]{Note4}%
  \BibitemOpen
  \bibinfo {note} {Note that one can also estimate the length $\delta l$ over
  which the oil invades the gap from the increase of the loss modulus $G''$.
  Indeed, since $G''$ corresponds to the energy dissipated per unit volume, at
  first order the relative variation of $G''$ follows: \protect \[\protect
  \frac {G''(\delta l)-G''(\delta l=0)}{G''(\delta l=0)} = 4\left (\protect
  \frac {\eta _o}{\eta _w}-1 \right )\protect \frac {\delta l}{R} \simeq
  4\protect \frac {\eta _o}{\eta _w}\protect \frac {\delta l}{R},\protect \] \\
  with \protect \[ G''\simeq \DOTSI \intop \ilimits@ _0^R \eta \protect
  \mathaccentV {dot}05F\gamma ^22\pi r e dr, \protect \] where $\protect
  \mathaccentV {dot}05F\gamma =\Omega r/e$ is the local shear rate, $\eta $ is
  the viscosity of the liquid phase (with $\eta =\eta _w$ for $r<R-\delta l$
  and $\eta =\eta _o$ for $r> R-\delta l$, where $\eta _o$ and $\eta _w$ stand
  respectively for the viscosity of the surrounding oil and that of water), $R$
  denotes the plate radius and $\Omega $ is the instantaneous angular velocity
  of the upper plate. From Fig.~\ref {fig.4b}(a), one infers that $\Delta
  G''/G''(\delta l=0) \simeq 0.7$ and with $\eta _o=60\eta _w$ one finds
  $\delta l\simeq 50\mu $m in excellent agreement with the average radial
  displacement of the oil/water interface derived from the spatiotemporal
  analysis of the snapshots pictured in Fig.~\ref {fig.4b}(b)-(e) (see
  Fig.~\ref {fig.sup1} and discussion in the appendix).}\BibitemShut {Stop}%
\bibitem [{\citenamefont {Ziane}\ \emph {et~al.}(2015)\citenamefont {Ziane},
  \citenamefont {Guirardel}, \citenamefont {Leng},\ and\ \citenamefont
  {Salmon}}]{Ziane:2015}%
  \BibitemOpen
  \bibfield  {author} {\bibinfo {author} {\bibfnamefont {N.}~\bibnamefont
  {Ziane}}, \bibinfo {author} {\bibfnamefont {M.}~\bibnamefont {Guirardel}},
  \bibinfo {author} {\bibfnamefont {J.}~\bibnamefont {Leng}}, \ and\ \bibinfo
  {author} {\bibfnamefont {J.-B.}\ \bibnamefont {Salmon}},\ }\bibfield  {title}
  {\enquote {\bibinfo {title} {Drying with no concentration gradient in large
  microfluidic droplets},}\ }\href@noop {} {\bibfield  {journal} {\bibinfo
  {journal} {Soft Matter}\ }\textbf {\bibinfo {volume} {11}},\ \bibinfo {pages}
  {3637--3642} (\bibinfo {year} {2015})}\BibitemShut {NoStop}%
\bibitem [{Note5()}]{Note5}%
  \BibitemOpen
  \bibinfo {note} {Note that the diffusion coefficient $D\simeq
  10^{-10}$~m$^2$.s$^{-1}$ of water in sunflower seed oil is an order of
  magnitude lower than the diffusivity of water in silicone oil reported in
  \cite {Ziane:2015} since the polar moieties of triglycerides strongly
  interact with water through hydrogen bonds, resulting in a lower mobility of
  dissolved water molecules \cite {Zieverink:2009}.}\BibitemShut {Stop}%
\bibitem [{\citenamefont {Zieverink}\ \emph {et~al.}(2009)\citenamefont
  {Zieverink}, \citenamefont {de~Rijke}, \citenamefont {de~Kruijf},\ and\
  \citenamefont {de~Kok}}]{Zieverink:2009}%
  \BibitemOpen
  \bibfield  {author} {\bibinfo {author} {\bibfnamefont {M.}~\bibnamefont
  {Zieverink}}, \bibinfo {author} {\bibfnamefont {E.}~\bibnamefont {de~Rijke}},
  \bibinfo {author} {\bibfnamefont {K.}~\bibnamefont {de~Kruijf}}, \ and\
  \bibinfo {author} {\bibfnamefont {P.}~\bibnamefont {de~Kok}},\ }\bibfield
  {title} {\enquote {\bibinfo {title} {Diffusivity and solubility of water in
  palm oil.}}\ }in\ \href@noop {} {\emph {\bibinfo {booktitle} {7th Euro Fed
  Lipid Congress, Poster (PHYS-004) Graz, Austria}}}\ (\bibinfo {year}
  {2009})\BibitemShut {NoStop}%
\bibitem [{\citenamefont {Medin}(1995)}]{Medin:1995}%
  \BibitemOpen
  \bibfield  {author} {\bibinfo {author} {\bibfnamefont {A.}~\bibnamefont
  {Medin}},\ }\emph {\bibinfo {title} {Studies on structure and properties of
  agarose}},\ \href@noop {} {Ph.D. thesis},\ \bibinfo  {school} {Uppsala}
  (\bibinfo {year} {1995})\BibitemShut {NoStop}%
\bibitem [{\citenamefont {Arevalo}\ \emph {et~al.}(2015)\citenamefont
  {Arevalo}, \citenamefont {Kumar}, \citenamefont {Urbach},\ and\ \citenamefont
  {Blair}}]{Arevalo:2015}%
  \BibitemOpen
  \bibfield  {author} {\bibinfo {author} {\bibfnamefont {R.}~\bibnamefont
  {Arevalo}}, \bibinfo {author} {\bibfnamefont {P.}~\bibnamefont {Kumar}},
  \bibinfo {author} {\bibfnamefont {J.}~\bibnamefont {Urbach}}, \ and\ \bibinfo
  {author} {\bibfnamefont {D.}~\bibnamefont {Blair}},\ }\bibfield  {title}
  {\enquote {\bibinfo {title} {Stress heterogeneities in sheared type-i
  collagen networks revealed by boundary stress microscopy},}\ }\href@noop {}
  {\bibfield  {journal} {\bibinfo  {journal} {PLoS ONE}\ }\textbf {\bibinfo
  {volume} {10}},\ \bibinfo {pages} {e0118021} (\bibinfo {year}
  {2015})}\BibitemShut {NoStop}%
\bibitem [{\citenamefont {Sonwai}\ and\ \citenamefont
  {Mackley}(2006)}]{Sonwai:2006}%
  \BibitemOpen
  \bibfield  {author} {\bibinfo {author} {\bibfnamefont {S.}~\bibnamefont
  {Sonwai}}\ and\ \bibinfo {author} {\bibfnamefont {M.}~\bibnamefont
  {Mackley}},\ }\bibfield  {title} {\enquote {\bibinfo {title} {The effect of
  shear on the crystallization of cocoa butter},}\ }\href@noop {} {\bibfield
  {journal} {\bibinfo  {journal} {Journal of the American Oil Chemists'
  Society}\ }\textbf {\bibinfo {volume} {83}},\ \bibinfo {pages} {583--586}
  (\bibinfo {year} {2006})}\BibitemShut {NoStop}%
\bibitem [{\citenamefont {Taylor}\ \emph {et~al.}(2009)\citenamefont {Taylor},
  \citenamefont {Damme}, \citenamefont {Johns},\ and\ \citenamefont {ans
  D.I.~Wilson}}]{Taylor:2009}%
  \BibitemOpen
  \bibfield  {author} {\bibinfo {author} {\bibfnamefont {J.}~\bibnamefont
  {Taylor}}, \bibinfo {author} {\bibfnamefont {I.~V.}\ \bibnamefont {Damme}},
  \bibinfo {author} {\bibfnamefont {M.}~\bibnamefont {Johns}}, \ and\ \bibinfo
  {author} {\bibfnamefont {A.~R.}\ \bibnamefont {ans D.I.~Wilson}},\ }\bibfield
   {title} {\enquote {\bibinfo {title} {Shear rheology of molten crumb
  chocolate},}\ }\href@noop {} {\bibfield  {journal} {\bibinfo  {journal}
  {Journal of Food Science}\ }\textbf {\bibinfo {volume} {74}},\ \bibinfo
  {pages} {55--61} (\bibinfo {year} {2009})}\BibitemShut {NoStop}%
\bibitem [{\citenamefont {Habouzit}(2012)}]{Habouzit:2012}%
  \BibitemOpen
  \bibfield  {author} {\bibinfo {author} {\bibfnamefont {D.}~\bibnamefont
  {Habouzit}},\ }\emph {\bibinfo {title} {Rh\'eophysique et stabilit\'e des
  mat\'eriaux h\'et\'erog\`enes solide/liquide \`a base de corps gras}},\
  \href@noop {} {Ph.D. thesis},\ \bibinfo  {school} {Universit\'e de Bordeaux
  I} (\bibinfo {year} {2012})\BibitemShut {NoStop}%
\bibitem [{\citenamefont {Peixinho}\ \emph {et~al.}(2010)\citenamefont
  {Peixinho}, \citenamefont {Karanjkar}, \citenamefont {Lee},\ and\
  \citenamefont {Morris}}]{Peixinho:2010}%
  \BibitemOpen
  \bibfield  {author} {\bibinfo {author} {\bibfnamefont {J.}~\bibnamefont
  {Peixinho}}, \bibinfo {author} {\bibfnamefont {P.}~\bibnamefont {Karanjkar}},
  \bibinfo {author} {\bibfnamefont {J.}~\bibnamefont {Lee}}, \ and\ \bibinfo
  {author} {\bibfnamefont {J.}~\bibnamefont {Morris}},\ }\bibfield  {title}
  {\enquote {\bibinfo {title} {Rheology of hydrate forming emulsions},}\
  }\href@noop {} {\bibfield  {journal} {\bibinfo  {journal} {Langmuir}\
  }\textbf {\bibinfo {volume} {26}},\ \bibinfo {pages} {11699--11704} (\bibinfo
  {year} {2010})}\BibitemShut {NoStop}%
\bibitem [{\citenamefont {Ahuja}, \citenamefont {Zylyftari},\ and\
  \citenamefont {Morris}(2015)}]{Ahuja:2015}%
  \BibitemOpen
  \bibfield  {author} {\bibinfo {author} {\bibfnamefont {A.}~\bibnamefont
  {Ahuja}}, \bibinfo {author} {\bibfnamefont {G.}~\bibnamefont {Zylyftari}}, \
  and\ \bibinfo {author} {\bibfnamefont {J.}~\bibnamefont {Morris}},\
  }\bibfield  {title} {\enquote {\bibinfo {title} {Yield stress measurements of
  cyclopentane hydrate slurry},}\ }\href@noop {} {\bibfield  {journal}
  {\bibinfo  {journal} {J. Non-Newtonian Fluid Mech.}\ }\textbf {\bibinfo
  {volume} {220}},\ \bibinfo {pages} {116--125} (\bibinfo {year}
  {2015})}\BibitemShut {NoStop}%
\bibitem [{\citenamefont {Chambon}\ and\ \citenamefont
  {Winter}(1987)}]{Chambon:1987}%
  \BibitemOpen
  \bibfield  {author} {\bibinfo {author} {\bibfnamefont {F.}~\bibnamefont
  {Chambon}}\ and\ \bibinfo {author} {\bibfnamefont {H.}~\bibnamefont
  {Winter}},\ }\bibfield  {title} {\enquote {\bibinfo {title} {Linear
  viscoelasticity at the gel point of a crosslinking pdms with imbalanced
  stoichiometry},}\ }\href@noop {} {\bibfield  {journal} {\bibinfo  {journal}
  {J. Rheol.}\ }\textbf {\bibinfo {volume} {31}},\ \bibinfo {pages} {683--697}
  (\bibinfo {year} {1987})}\BibitemShut {NoStop}%
\end{thebibliography}

%

\end{document}